\documentclass[fleqn,usenatbib]{mnras}

\usepackage{newtxtext,newtxmath}

\usepackage[T1]{fontenc}

\DeclareRobustCommand{\VAN}[3]{#2}
\let\VANthebibliography\thebibliography
\def\thebibliography{\DeclareRobustCommand{\VAN}[3]{##3}\VANthebibliography}

\usepackage{graphicx}	
\usepackage{amsmath}	
\usepackage{amssymb}	
\usepackage{color}
\usepackage{mathabx}
\usepackage{epstopdf}
\usepackage[dvipsnames]{xcolor}
\usepackage{ulem}

\renewcommand{\textcolor}[2]{#2}{}

\title[MRI-active inner disc]{MRI-active inner regions of protoplanetary discs. I. A detailed model of disc structure}

\author[]{Marija R. Jankovic$^{1}$\thanks{E-mail: mj577@cam.ac.uk},
James E. Owen$^{2}$,
Subhanjoy Mohanty$^{2}$
and Jonathan C. Tan$^{3, 4}$
\\
$^{1}$Institute of Astronomy, University of Cambridge, Madingley Road, Cambridge CB3 0HA, UK\\
$^{2}$Astrophysics Group, Imperial College London, Blackett Laboratory, Prince Consort Road, London SW7 2AZ, UK\\
$^{3}$Dept. of Astronomy, University of Virginia, Charlottesville, Virginia 22904, USA\\
$^{4}$Dept. of Space, Earth \& Environment, Chalmers University of Technology, Gothenburg, Sweden
}

\date{Accepted XXX. Received YYY; in original form ZZZ}

\pubyear{2020}

\begin{document}
\label{firstpage}
\pagerange{\pageref{firstpage}--\pageref{lastpage}}
\maketitle

\begin{abstract}
Short-period super-Earth-sized planets are common. Explaining how they form near their present orbits requires understanding the structure of the inner regions of protoplanetary discs. Previous \textcolor{red}{studies} have argued that the hot inner protoplanetary disc is unstable to the magneto-rotational instability (MRI) due to thermal ionization of potassium, \textcolor{red}{and that a local gas pressure maximum forms at the outer edge of this MRI-active zone}. Here we present a steady-state model \textcolor{red}{for inner discs} accreting viscously, primarily due to the MRI. \textcolor{red}{The structure and MRI-viscosity of the inner disc are fully coupled in our model; moreover, we account for many processes omitted in previous such models, including disc heating by both} accretion and stellar irradiation, vertical energy transport, \textcolor{red}{realistic} dust opacities, dust effects on disc ionization and non-thermal sources of ionization. \textcolor{red}{
\textcolor{red}{For a disc around a solar-mass star with a standard gas accretion rate ($\dot{M}$$\sim$$10^{-8}$\,M$_\odot$\,yr$^{-1}$) and small dust grains, we find that} the inner disc is optically thick, and the accretion heat is primarily released near the midplane. As a result, \textcolor{red}{both} the disc midplane temperature and the location of the pressure maximum are only \textcolor{red}{marginally} affected by stellar irradiation, and the \textcolor{red}{inner} disc is also convectively unstable. As previously suggested, the inner disc is primarily ionized through thermionic and potassium ion emission from dust grains, \textcolor{red}{which, at high temperatures,} counteract adsorption of free charges onto grains. Our results show \textcolor{red}{that} the location of the pressure maximum is determined by the threshold temperature above which thermionic and ion emission become efficient.}
\end{abstract}




\section{Introduction}
Close-in super-Earths -- planets with radii 1--4\,R$_\oplus$ and orbital periods shorter than $\sim$100 days -- are common around solar-type and lower mass stars \citep{Fressin2013, Dressing2013, Dressing2015, Mulders2018, Hsu2019,Zink2019}. Yet how these planets form is still an open problem. One theory posits that they are born further away from the star, and subsequently migrate through the protoplanetary disc to their present orbits \citep{Terquem2007, Ogihara2009, McNeil2010, Cossou2014, Izidoro2017, Izidoro2019, Bitsch2019}. However, this hypothesis predicts that the planets should be water-rich, in contrast to the water-poor composition inferred from their observed radius distribution combined with atmospheric evolution models \citep{Owen2017,VanEylen2018,Wu2019}. 

An alternative proposal is that these super-Earths form at or near their present orbits, out of solid material which migrates to the inner disc prior to planet formation. These solids are expected to arrive in the form of pebbles, which radially drift inwards from the outer disc due to gas drag \citep{Hansen2013, Boley2013, Chatterjee2014, Hu2018,Jankovic2019}. However, these pebbles need to be trapped, i.e., their radial drift must be halted, in order for them to coalesce into planets instead of drifting into the star. Such trapping may occur as follows. 

In the inner protoplanetary disc, at the short orbital periods at which close-in super-Earths are observed, the gaseous disc is thought to accrete via the magneto-rotational instability \citep[MRI;][]{Balbus1991}. The MRI leads to turbulence in the disc, which drives viscous accretion, but its magnitude is sensitive to the disc's ionization state. The MRI is expected to be efficient in the hot, innermost parts of the disc (where it can be activated by thermal ionization of trace alkali elements, at temperatures $\gtrsim$1000\,K), and largely suppressed in cold regions further away from the star \citep[the latter being known as the MRI dead zone;][]{Gammie1996}. At the transition between the two regions, a local gas pressure maximum forms when the disc is in steady state, and this pressure maximum may trap the pebbles that drift in from the outer disc \citep{Chatterjee2014}. Concurrently, the MRI-driven turbulence in the inner disk (interior to the pressure maximum) limits the grain size there by inducing high collisional velocities between grains, causing them to fragment. A decrease in the size of grains reduces their radial drift, providing an alternative way to accumulate the grains arriving from the outer disc \citep{Jankovic2019}. In general, therefore, the structure of the inner disc, undergoing MRI-induced accretion, is likely to play a key role in the formation of close-in super-Earths.

In previous work \citep{Mohanty2018}, we presented a model of the inner disc in which the disc structure and accretion due to the MRI are treated self-consistently. We obtained an inner disc structure in line with the expectations sketched above, and inferred a location for the pressure maximum consistent with the orbital distances of close-in super-Earths. However, this model includes a number of simplifying assumptions about the disc's physical and chemical structure. First, the disc is considered to be vertically isothermal, and second, heating by stellar irradiation is neglected. In reality, the disc temperature will vary vertically, with a profile that is particularly non-trivial when both accretion heating and stellar irradiation are accounted for \citep[e.g.][]{DAlessio1998}. Third, the model assumes a constant dust opacity, when in fact the latter depends on the temperature and on grain properties and abundance.

Fourth, \citet{Mohanty2018} assumed that the only source of ionization in the disc is thermal (collisional) ionization of potassium, and only considered gas-phase interactions. Potassium is indeed a good representative of thermally-ionized species in the inner disc due to its low ionization potential and high abundance \citep{Desch2015}. However, ions and free electrons are also adsorbed onto the surfaces of dust grains in the disc, where they quickly recombine. Dust can thus reduce the disc ionization level and suppress the MRI \citep{Sano2000, Ilgner2006, Wardle2007, Salmeron2008, Bai2009, Mohanty2013}. Conversely, hot grains ($\gtrsim 500$\,K) can also emit electrons and ions into the gas phase (for electrons, this process is known as thermionic emission). Such temperatures are easily attainable in the inner disc, and thermionic and ion emission \textcolor{red}{have been hypothesized} to be important sources of ionization there \citep{Desch2015}. Neither of these grain effects are treated by \citet{Mohanty2018}. Finally, while \citet{Mohanty2018} showed a posteriori that, in the MRI-accreting inner disc, X-ray ionization of molecular hydrogen may be competitive with thermal ionization of potassium in supplying free electrons (due to the low gas surface density in this region), they did not actually include X-ray ionization in their model. 


In this paper we present a new model of an MRI-accreting, steady-state inner disc that addresses each of the shortcomings of the \citet{Mohanty2018} model listed above. The vertical structure in our model is calculated self-consistently from viscous dissipation (due to the MRI-induced viscosity), stellar irradiation, and radiative and convective cooling, with realistic opacities due to dust grains. Ionization in the disc is determined by thermal ionization of potassium, thermionic and ion emission from dust grains, and ionization of molecular hydrogen by stellar X-rays, cosmic rays and radionuclides. In \S\ref{sec:methods} we detail all the components of our model and the methods used to find self-consistent steady-state solutions for the disc structure. We present our results in \S\ref{sec:results}, discuss the relative importance of the various physical and chemical processes in \S\ref{sec:discussion}, and summarise our findings in \S\ref{sec:summary}. 

\textcolor{red}{The aim of this paper is two-fold: to present the methodology of our calculations, and to discuss the detailed physics of the inner disc in the context of fiducial disc and stellar parameters. In a companion paper (Paper II: Jankovic et al. {\it in prep.}), we investigate how the inner disc structure varies as a function of these parameters, and discuss the implications for the formation of super-Earths in the inner disc.}

\section{Methods} \label{sec:methods}
We consider a disc that is viscously accreting and in steady-state, i.e., has a constant mass accretion rate. Our model of the disc structure is described in \S\ref{sec:disc_model}. The disc structure depends on the disc's radiative properties, i.e. opacities, and on the viscosity. Our calculation of opacities is summarised in \S\ref{sec:opacities}, and the prescription for MRI-driven viscosity given in \S\ref{sec:mri_alpha}. The latter viscosity is a function of the disc's ionization state, calculated using a chemical network described in \S\ref{sec:ionization}. The disc structure, opacities, ionization and viscosity are calculated self-consistently at every point in the disc, using numerical methods supplied in \S\ref{sec:numerical_methods}.

The key parameters of our model are the steady-state gas accretion rate $\dot{M}$ through the disc, stellar mass $M_*$, stellar radius $R_*$, stellar effective temperature $T_*$, and value of the viscosity in the absence of the MRI (the dead-zone viscosity). Additionally, the disc opacities and ionization state, and thus the disc structure, depend on the properties of the dust; most importantly, on the dust-to-gas ratio $f_{\rm dg}$ and the maximum dust grain size $a_{\rm max}$.

\textcolor{red}{We assume that the disc only accretes viscously, due to the MRI. Disc accretion may \textcolor{red}{additionally} be partially driven by magnetic winds, which could also affect the inner disc structure \citep[e.g., ][]{Suzuki2016}. Therefore, inclusion of wind-driven accretion is an important issue\textcolor{red}{; however, it is one which we do not tackle here}.}

\subsection{Disc model} \label{sec:disc_model}
Our disc model largely follows the work of \citet{DAlessio1998, DAlessio1999}. We consider a thin, axisymmetric, Keplerian, steady-state disc that is viscously accreting. We assume that the disc is in vertical hydrostatic equilibrium, heated by viscous dissipation and stellar irradiation, and that energy is transported by radiation and convection. Since the disc is vertically thin, we neglect energy transport in the radial direction. Furthermore, at a given disc radius, our viscosity depends only on local conditions and the vertical mass column (see \S\ref{sec:mri_alpha}). As such, the disc structures at different radii are only coupled by the stellar irradiation, as it penetrates the disc along the line-of-sight to the central star. 

\subsubsection{Hydrostatic equilibrium}
In a thin Keplerian disc in vertical hydrostatic equilibrium, the gas pressure profile at any given radius follows from
\begin{equation} \label{eq:hydrostatic_equilibrium}
  \frac{d P}{d z} = - \rho \Omega^2 z ,
\end{equation}
where $P$ is the gas pressure, $\rho$ the gas volume density, $\Omega$ the Keplerian angular velocity, and $z$ the height above the disc midplane. We adopt the ideal gas law.

\subsubsection{Viscous heating and stellar irradiation} \label{sec:heating_terms}
The disc is heated by the viscosity that drives the accretion. The viscosity $\nu$ is parametrized by $\nu = \alpha c_{\rm s}^2/\Omega$, where $\alpha$ is the viscosity parameter and $c_{\rm s}$ the isothermal sound speed \citep{Shakura1973}. The local viscous dissipation rate at any location in the disc is given by
\begin{equation}
  \Gamma_{\rm acc} = \frac{9}{4} \alpha P \Omega .
\end{equation}
The flux generated by viscous dissipation that is radiated through one side of the disc at any disc radius $r$ is
\begin{equation} \label{eq:total_viscous_heating}
  F_{\rm acc} = \frac{3}{8\pi} \dot{M} f_r \Omega^2 ,
\end{equation}
where $f_r=1-\sqrt{R_{\rm in}/r}$ comes from the thin boundary-layer condition at the inner edge of the disc, $R_{\rm in}=R_*$ is the radius of the inner edge of the disc, and we are assuming a zero-torque inner boundary condition \citep[e.g.][]{Frank2002}. 

We also consider heating due to stellar irradiation. Stellar flux propagates spherically outwards from the star, and the resultant heating at any disc location depends on the attenuation of this flux along the line-of-sight to the star. Accounting for this, however, while simultaneously neglecting scattering of starlight and radial energy transport within the disc (as we do in our 1+1D model here), leads to multiple equilibrium solutions or none at all \citep[e.g.][]{Chiang2001}. Such behaviour does not appear in 2D disc models \citep[e.g.][]{Dullemond2002}, which are nevertheless too complex for our purposes here. Instead, we treat the irradiation heating at each disc radius in isolation, by considering heating only due to the stellar flux that impinges on the disc surface at that radius (at some grazing angle $\phi$ calculated self-consistently; see further below), and then propagates vertically towards the midplane \citep{Calvet1992,Chiang1997}. In this framework, the attenuation, i.e. the optical depth to the stellar flux, 
is approximated as $\tau_{\rm irr} \approx \tau_{\rm{irr},z}/\mu$, where $\tau_{\rm{irr},z}$ is the optical depth in the vertical direction, and $\mu\equiv\textrm{sin} \phi$. Local heating due to stellar irradiation is then given by
\begin{equation}
  \Gamma_{\rm irr} = \kappa_{\rm P}^* \rho \frac{F_{\rm irr}}{\mu} e^{-\tau_{\rm{irr},z}/\mu} ,
\end{equation}
where $\kappa_{\rm P}^*$ is the disc Planck opacity to stellar irradiation (see section \ref{sec:opacities}) and $F_{\rm irr}$ is the total incident stellar flux at the specified disc radius. In an optically-thick disc, the latter is given by 
\begin{equation} \label{eq:total_irradiation_heating}
  F_{\rm irr} = \sigma_{SB} T_*^4 \left( \frac{R_*}{s} \right)^2 \mu ,
\end{equation}
where $s$ is the distance to the star (spherical radius) from the disc surface at the specified radius.\footnote{When calculating the factor $f_r$ in eq. (\ref{eq:total_viscous_heating}) we have assumed that the inner disc edge is at the stellar radius, which would imply that the disc surface can only see a half of the stellar disc. If the disc surface can only see a half of the stellar disc, the expression for the total absorbed stellar flux, eq. (\ref{eq:total_irradiation_heating}), should include an additional factor of $1/2$ \citep[e.g.][]{Chiang1997}. However, in a more realistic case of the disc being truncated at several stellar radii, the entire stellar disc should be visible \citep[see e.g.][]{Estrada2016}, and so we do not include this additional factor.}

Finally, the grazing angle $\phi$ is given by
\begin{equation} \label{eq:graz_angle}
	\phi = \textrm{sin}^{-1} \frac{4}{3\pi} \frac{R_*}{r} + \textrm{tan}^{-1} \frac{d \textrm{log} z_{\rm irr}}{d \textrm{log} r} \frac{z_{\rm irr}}{r} - \textrm{tan}^{-1} \frac{z_{\rm irr}}{r} .
\end{equation}
Here the first term is the value of $\phi$ for a flat disc and comes from the finite size of the stellar disc, and the other two terms are due to disc flaring. $z_{\rm irr}(r)$ is the height above the disc midplane at which the stellar flux is absorbed; we take $z_{\rm irr}(r)$ to be the height where the optical depth to the stellar irradiation is $\tau_{\rm irr} = 2/3$. Specifically, in calculating the height $z_{\rm irr}$, the optical depth $\tau_{\rm irr}$ is obtained by integrating along the spherical radius $s$ to the star, as opposed to the approximation used in the local heating term. We determine $z_{\rm irr}$ and $\phi$ self-consistently with the disc structure following the procedure by \citet{DAlessio1999}, as outlined in section \ref{sec:numerical_methods}.

\subsubsection{Radiative energy transport} \label{sec:radiative_transport}
In a thin disc, the optical depth to the disc's own radiation is much smaller in the vertical direction than in the radial. Thus we expect radiative energy transport to be primarily vertical, and that is the only direction we consider.  

The frequency-integrated moments of the radiative transfer equation in the Eddington approximation (i.e. assuming that the radiation is isotropic, as is valid in the optically thick regime) and the equation for energy balance are then
\begin{align}
  \frac{dF}{dz} &= \Gamma_{\rm acc} + \Gamma_{\rm irr} , \label{eq:rt_1} \\ 
  \frac{dJ}{dz} &= - \frac{3 \rho \kappa_{\rm R}}{4 \pi} F , \label{eq:rt_2} \\ 
  4 \rho \kappa_{\rm P} (\sigma_{\rm SB} T^4 - \pi J) &= \Gamma_{\rm acc} + \Gamma_{\rm irr} , \label{eq:rt_3} 
\end{align}
where $F$ and $J$ are the radiative flux and mean intensity respectively. We have also assumed here that the $J$ and $F$ weighted opacities can be approximated by the Planck mean opacity $\kappa_{\rm P}$ and the Rosseland mean opacity $\kappa_{\rm R}$ respectively \citep[following e.g.][see \S\ref{sec:opacities}]{Hubeny1990}.

Together with the ideal gas equation of state, equations (\ref{eq:hydrostatic_equilibrium}, \ref{eq:rt_1}-\ref{eq:rt_3}) form a closed set in $P$, $F$, $J$ and $T$. Together with appropriate boundary conditions, they determine the disc vertical structure. One boundary condition is imposed at the disc midplane, where the flux $F(0)=0$ by symmetry. The remaining boundary conditions are supplied at the disc surface, at height $z_{\rm surf}$ above the midplane. The boundary condition for the flux $F$ there is obtained by integrating eq. (\ref{eq:rt_1}) from $z=0$ to $z_{\rm surf}$, from which it follows that $F(z_{\rm surf}) = F_{\rm acc} + F_{\rm irr}$. The boundary condition for the mean intensity $J$ is given by $J(z_{\rm surf}) = \frac{1}{2\pi} F(z_{\rm surf})$. Finally, we assume that the gas pressure at the top of the disc has a small constant value, $P(z_{\rm surf}) = 10^{-10} \textrm{\,dyn\,cm}^{-2}$, which is another boundary condition. The precise value of $P(z_{\rm surf})$ is arbitrary, but does not affect our results as long as it is sufficiently small. Overall, then, we have four boundary conditions on three differential equations. Note that the temperature $T(z_{\rm surf})$ at the disc surface follows from the algebraic eq. (\ref{eq:rt_3}), the boundary conditions on $J$ and $P$, and the ideal gas law.


\subsubsection{Energy transport by convection} \label{sec:convective_transport}
If radiative energy transport yields a thermal structure such that the gradient $\nabla = \frac{d \textrm{ln} T}{d \textrm{ln} P}$ is greater than the adiabatic gradient $\nabla_{\rm ad} = (\gamma-1)/\gamma$, then the gas is unstable to convection. In disc regions where this is the case, we assume that energy transport by convection is efficient and the gas is vertically isentropic, so that $\nabla = \nabla_{\rm ad}$ at such locations \citep[e.g.][]{Shu1992, Rafikov2007,Garaud2007}. We adopt $\gamma = 1.4$, valid for an H$_2$ dominated disc. Equations (\ref{eq:rt_2}) and (\ref{eq:rt_3}) are then replaced by 
\begin{align} \label{eq:convection}
	\frac{d T}{d z} &= - \nabla_{\rm ad} \frac{T}{P} \rho(T, P) \Omega^2 z = - \nabla_{\rm ad} \frac{\mu m_{\rm H}}{k_{\rm B}} \Omega^2 z.
\end{align}

\subsection{Opacities} \label{sec:opacities}
Radiative transport is controlled by the Rosseland-mean opacity $\kappa_{\rm R}$ in optically-thick regions (eq. \ref{eq:rt_2}), and by the Planck-mean opacity $\kappa_{\rm P}$ in optically-thin regions (eq. \ref{eq:rt_3}). Additionally, the absorption coefficient for the stellar flux is a Planck-mean opacity $\kappa_{\rm P}^*$ at the stellar effective temperature. We assume that the only source of these opacities are dust grains. Gas opacities are important in the very innermost regions of discs, where most dust species have sublimated. However, these regions are not of particular significance for the early stages of planet formation that we are interested in, as dust is required to form solid planet cores. Beyond the silicate sublimation line, gas opacities may still be important in hot, optically thin regions \citep{Malygin2014}. However, as we find, the inner disc is significantly optically thick and including gas opacities would only alter the structure of the hot disc atmosphere. Therefore, we completely ignore the contribution of gas to the opacities.

Furthermore, we assume that the only dust species present are silicate grains. Other species that can be comparable in abundance to silicates are water ice and carbonaceous grains \citep[e.g. organics,][]{Pollack1994}. However, due to their low sublimation temperatures, neither water ice nor carbonaceous grains are expected in the hot MRI-active regions, and we generally limit our calculations to the inner 1\,AU of the disc. 

To calculate the opacities, we adopt the optical constants of ``astronomical silicates'' from \citet{Draine2003}. We use the \textsc{miescat} module of the python wrapper \textsc{radmc3dPy} for \textsc{RADMC3D} \citep{Dullemond2012} to obtain dust absorption and scattering coefficients as functions of radiation wavelength and grain size. For each grain size $a$, the coefficients are averaged within a size bin of width $\Delta \textrm{ln} a=0.02$. Next, we assume that the grain bulk density is $\rho_{\rm gr}=3.3$\,g\,cm$^{-3}$, and that the grain sizes are described by the MRN distribution, wherein the number density of grains in the size range $[a,a+da]$ is given by $n(a) da$\,$\propto$\,$a^{-q} da$ \citep{Mathis1977}, with $q=3.5$. We adopt a minimum grain size of $a_{\rm min}=0.1$\,$\mu$m, and a variable maximum grain size $a_{\rm max}$. The size-dependent absorption and scattering coefficients are then weighted by grain mass and averaged over the grain size distribution. Finally, the absorption coefficient is integrated over frequency to obtain the Planck-mean opacity $\kappa_{\rm P}(T)$, and the total extinction coefficient yields the Rosseland-mean opacity $\kappa_{\rm R}(T)$. Following \citet{DAlessio1998}, we calculate the mean absorption coefficient for the stellar flux as a frequency-integrated absorption coefficient weighted by the Planck function at the stellar effective temperature: $\kappa_{\rm P}^*\equiv\kappa_{\rm P}(T_*)$.

\begin{figure}
    \centering
    \includegraphics[width=\columnwidth]{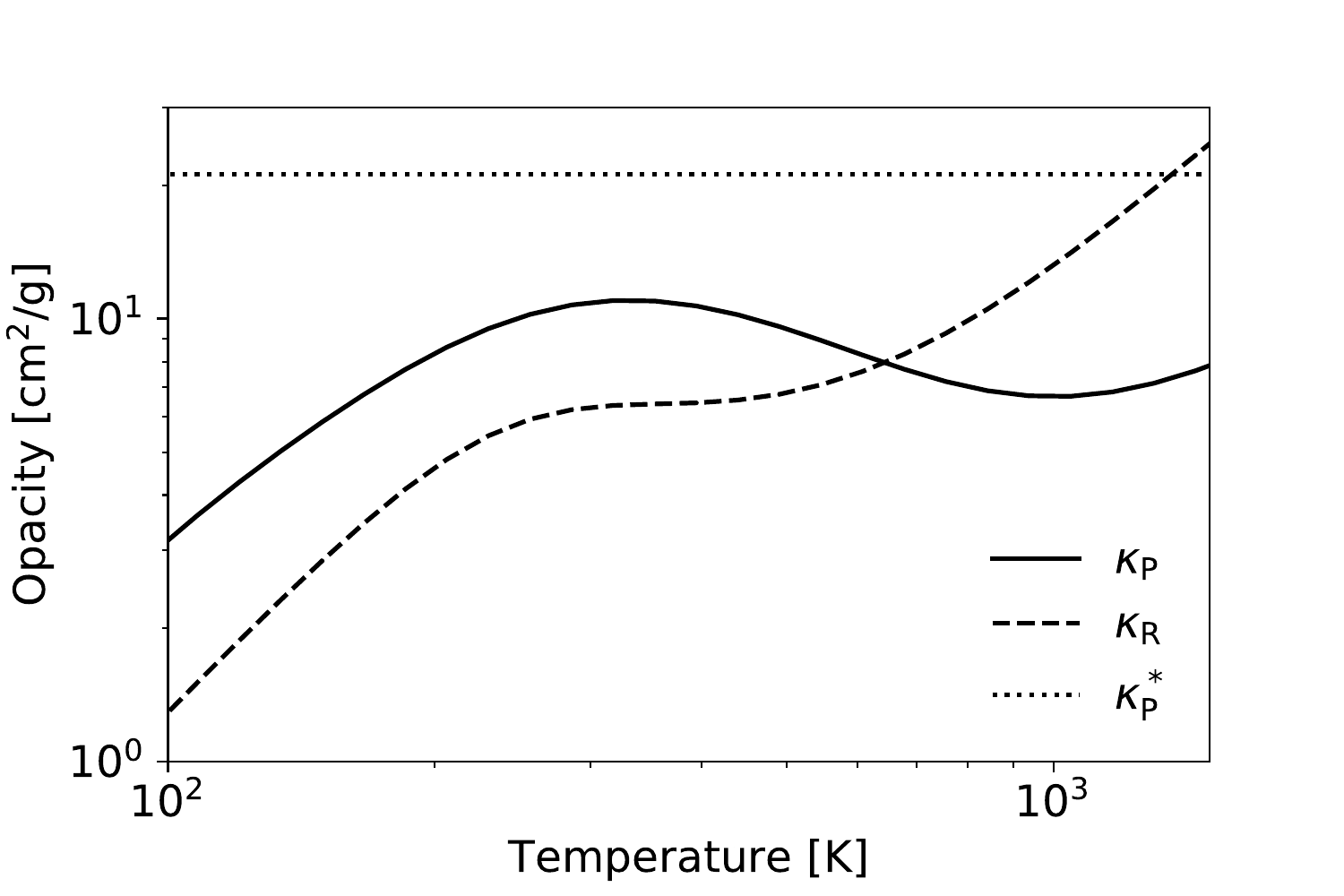}
    \caption{Planck-mean opacity $\kappa_{\rm P}$, Rosseland-mean opacity $\kappa_{\rm R}$, and Planck-mean opacity at the stellar effective temperature $\kappa_{\rm P}^*$, as functions of disc temperature, assuming dust-to-gas ratio $f_{\rm dg}=10^{-2}$ and a maximum dust grain size $a_{\rm max}=1$\,$\mu$m.}
    \label{fig:opacities}
\end{figure}

Figure \ref{fig:opacities} shows the opacities per unit mass of gas, assuming a dust-to-gas ratio of $f_{\rm dg}=10^{-2}$, a maximum grain size of $a_{\rm max}=1$\,$\mu$m, and $T_*=4400$\,K. The Planck-mean opacity at the stellar effective temperature $\kappa_{\rm P}^*$ is a constant, since the wavelength-dependent dust absorption coefficient does not depend on the local temperature. The Planck-mean opacity $\kappa_{\rm P}$ is in general expected to increase with increasing temperature. This is because the wavelength at which the Planck function peaks is inversely proportional to the temperature, and for grains smaller than the wavelength of peak emission (and small grains contribute to the opacities most) absorption is expected to increase with decreasing wavelength. However, due to the silicate absorption feature at 10\,$\mu$m, $\kappa_{\rm P}$ decreases with temperature in the range $\sim$\,500\,K -- 1000\,K. 


\subsection{Viscosity} \label{sec:mri_alpha}
Our model of the MRI-driven viscosity closely follows that of \citet{Mohanty2018} \citep[see also][]{Bai2011b}. Here we only summarize the main points. A well-ionized circumstellar disc which follows the laws of ideal magnetohydrodynamics (MHD) is susceptible to the MRI \citep{Balbus1991}. The MRI leads to turbulence, producing an accretion stress and acting as a source of viscosity. The resulting Shakura-Sunyaev viscosity parameter $\alpha$ is
\begin{equation}
\label{eq:alpha_beta}
	\alpha_{\rm AZ} = \frac{1}{3\beta},
\end{equation}
where the subscript `AZ' denotes an MRI-active zone, $\beta \equiv P/P_B$ is the plasma parameter, and $P_B = B^2/8\pi$ is the magnetic field pressure \citep{Sano2004}\footnote{The numerical factor is indeed $1/3$ for the Shakura-Sunyaev $\alpha$ parameter in eq. (\ref{eq:alpha_beta}); see discussion in Appendix B of \citet{Mohanty2018}.}. 

However, even in the inner regions of protoplanetary discs, non-ideal MHD effects can quench the MRI, leading to so-called dead zones. The non-ideal effects considered here are Ohmic and ambipolar diffusion. Ohmic diffusion will \textit{not} suppress the MRI if \citep{Sano2002}
\begin{equation} \label{eq:ohm_crit}
	\Lambda = \frac{v_{{\mathcal A}z}^2}{\eta_{\rm O} \Omega} > 1 ,
\end{equation}
where $\Lambda$ is the Ohmic Elsasser number, $v_{{\mathcal A}z} \equiv B_z/\sqrt{4\pi\rho}$ is the vertical component of the local Alfven velocity and $\eta_{\rm O}$ is the Ohmic resistivity. Here we utilize the relationship between the strength of the vertical component of the magnetic field, $B_z$, and the strength of the r.m.s. field, $B$: $B_z^2 \sim B^2 / 25$ \citep{Sano2004}. We also assume that $B$ is vertically constant. Our method of determining the value of $B$ is described in section \ref{sec:numerical_methods}.

Similarly, the ambipolar Elsasser number is defined by
\begin{equation}
    Am = \frac{v_{\mathcal A}^2}{\eta_{\rm A} \Omega} ,
\end{equation}
where $\eta_{\rm A}$ is the ambipolar magnetic resistivity. However, in the strong-coupling limit, valid in protoplanetary discs, the MRI can be active even if $Am < 1$, as long as the magnetic field is sufficiently weak \citep{Bai2011a}. The criterion for active MRI in the presence of ambipolar diffusion is
\begin{equation} \label{eq:amb_crit}
	\beta/\beta_{\rm min} > 1,
\end{equation}
where the minimum value of $\beta$ neccessary to sustain the MRI is a function of the ambipolar Elsasser number:
\begin{equation}
    \beta_{\rm min}(Am) = \left[ \left( \frac{50}{Am^{1.2}} \right)^2 + \left( \frac{8}{Am^{0.3}} + 1 \right)^2 \right]^{1/2} .
\end{equation}

Thus, whether the MRI is active or not depends on the magnetic resistivities, $\eta_{\rm O}$ and $\eta_{\rm A}$, as well as on the magnetic field strength $B$. The magnetic resistivities, calculated following \citet{Wardle1999}, express the coupling between the gas and the magnetic field, which is principally determined by the degree of ionization of the gas.

If either of the two criteria given by eqns. (\ref{eq:ohm_crit}) and (\ref{eq:amb_crit}) is not fulfilled, the MRI is not active. In such MRI-dead zones, we assume there is a small residual viscosity $\alpha_{\rm DZ}$, driven either by propagation of turbulence from the MRI-active zone, or by hydrodynamic instabilities \citep[for more discussion, see][]{Mohanty2018}. 
In this work, we also impose a smooth transition between the active and the dead zones, necessary to ensure numerical stability in the integration of the equations of disc structure \citep[such smoothing was not employed by][]{Mohanty2018}. To cover both non-ideal effects that lead to dead zones, we define $C \equiv \textrm{min} (\Lambda, \beta/\beta_{\rm min})$. Then, at any location in the disc \textcolor{red}{where} $|C-1| < 0.5$ \textcolor{red}{(i.e., in the vicinity of the switch from active to dead)}, we adopt \textcolor{red}{a smoothed $\alpha$ given by}
\begin{equation}
	\alpha = \alpha_{\rm DZ} + \frac{\alpha_{\rm AZ}-\alpha_{\rm DZ}}{1 + \exp\left(-\frac{C-1}{\Delta}\right)} ,
\end{equation}
where $\Delta = 10^{-2}$. 

\subsection{Ionization} \label{sec:ionization}
\subsubsection{Chemical network}
We implement the same simple chemical network adopted by \citet{Desch2015}. This network tracks the number densities of five species: free electrons ($n_{\rm e}$), potassium ions ($n_{\rm K^+}$), neutral potassium atoms ($n_{\rm K^0}$), potassium atoms adsorbed (condensed) onto dust grains ($n_{\rm K, cond}$), and atomic ions ($n_{\rm i}$; i.e., ions of atomic species other than potassium; see below). 

In the gas-phase, potassium atoms can be thermally ionized: collisions of neutral potassium atoms with H$_2$ molecules produce potassium ions and free electrons. The potassium ions and free electrons can \textcolor{red}{also recombine in the gas phase,} either radiatively or via three-body collisions with H$_2$ molecules (the latter process dominates at the high densities prevalent in the inner disc). 

Furthermore, non-thermal sources (we consider stellar X-rays, cosmic rays and radionuclides; see section \ref{Hionizationrate}) can ionize H$_2$ \citep{Glassgold1997,Ercolano2013}. The charge is quickly transferred from the ionized hydrogen to other abundant gas species through collisions, producing molecular and atomic ions (e.g. HCO$^+$, Mg$^+$). Notably, in application to the MRI, the exact composition of the gas that this leads to is unimportant in the presence of dust, and simple chemical networks reproduce the gas ionization levels well \citep{Ilgner2006}. Thus, it is assumed that the ionization of molecular hydrogen \textcolor{red}{at a rate $\zeta$} directly produces atomic ions and free electrons at a \textcolor{red}{volumetric} rate $\zeta n_{\rm H_2}$. The atomic ion species in this chemical network may thus be understood as a representative of the various chemical species abundant in the gas-phase, whose mass is taken to be that of magnesium. It is assumed that the number density of molecular hydrogen is constant, which is valid for low ionization rates. Just as potassium, the atomic ions also recombine radiatively and in three-body recombinations.

Importantly, all gas-phase species \textcolor{red}{(electrons, ions and neutral atoms)} collide with and are adsorbed onto dust grains at a rate
\begin{equation}
	\mathcal{R}_{j\textrm{, coll}} = n_j n_{\rm gr} \pi a_{\rm gr}^2 \left( \frac{8k_{\rm B}T}{\pi m_j} \right)^{1/2} \tilde{J}_j S_j ,
\end{equation}
where $n_j$ is the number density of the gas-phase species, $n_{\rm gr}$ is the number density of the grains, $a_{\rm gr}$ is the grain size, $m_j$ is the gas-phase species mass, $\tilde{J}_j$ is the modification of the collisional cross-sections for charged species due to dust grain charge \citep{Draine1987}, and $S_j$ is the sticking coefficient. It is assumed that all grains have the same charge; this is valid since the dispersion in the distribution of charge states is generally found to be small \citep{Draine1987}. It is further assumed that potassium ions rapidly recombine on the grain surface to form condensed potassium atoms. Thus the ions are effectively destroyed upon adsorption.

At high temperatures electrons on the dust grains have a finite probability of leaving the grain, producing so-called thermionic emission. The emission depends on the energy required for the electron to escape the grain. For a neutral grain this is the work function $W$, a property of the material out of which the grains are made. The rate at which free electrons are produced through thermionic emission is
\begin{equation}
	\mathcal{R}_{\rm therm} = n_{\rm gr} 4\pi a_{\rm gr}^2 \lambda_{\rm R} \frac{4\pi m_{\rm e} (k_{\rm B} T)^2}{h^3} \,{\rm exp}\left({-\frac{W_{\rm eff}}{k_{\rm B} T}}\right) ,
\end{equation}
where
\begin{equation}
	W_{\rm eff} = W + \frac{Z e^2}{a_{\rm gr}} \label{eqn:Weff}
\end{equation}
is the effective work function due to grain charge $Z e$.

Potassium atoms will also evaporate from the grains only at high temperatures. The vaporization rate of condensed potassium atoms is given by
\begin{equation}
	\mathcal{R}_{\rm K, evap} = n_{\rm K, cond} \nu\, {\rm exp}\left({-\frac{E_{\rm a}}{k_{\rm B}T}}\right) ,
\end{equation}
where $\nu$ is the vibration frequency of potassium atoms on the dust grain surface lattice, and $E_{\rm a}=3.26$\,eV is the binding energy, whose value is chosen to reproduce the condensation temperature of potassium \citep[1006\,K,][]{Lodders2003}. These potassium atoms may be emitted into the gas phase as both neutral atoms and ions, contributing further to the gas' ionization state. The ratio of ions to neutrals among the emitted particles is given by
\begin{equation}
\label{eq:pot_ion_atom_ratio}
	\frac{n_{\rm K}^+}{n_{\rm K^0}} = \frac{g_+}{g_0}\, {\rm exp}\left({+\frac{W_{\rm eff} - \rm{IP}}{k_{\rm B} T}}\right) ,
\end{equation}
where $\frac{g_+}{g_0}$ is the ratio of statistical weights of the ionized and neutral state of potassium, and $\rm{IP}$ the ionization potential of potassium. The fraction of all emitted particles that leave the grain as ions is then given by 
\begin{equation}
f_+ = \frac{n_{\rm K}^+/n_{\rm K^0}}{1+n_{\rm K}^+/n_{\rm K^0}},
\end{equation}
so the rate at which potassium ions evaporate from the grains is given by $\mathcal{R}_{\rm K, evap} f_+$.

\textcolor{red}{It is important to note here that there are two sources of the neutral potassium atoms condensed on grains, whose evaporation is described above: first, gas-phase potassium ions that are adsorbed onto grain surfaces and recombine there into neutral atoms; and second, gas-phase neutral potassium atoms that are adsorbed directly onto the grains. The reionisation and subsequent ejection from grains of the former simply returns ions to the gas-phase that were originally adsorbed from it, and thus clearly cannot increase the ion fraction in the gas-phase beyond what it would be in the absence of grains. The ionisation on the grain surface, and ejection as ions, of particles that were originally {\it neutral} in the gas-phase, however, {\it can} increase the gas-phase ion fraction beyond what it would be without grains; it is this channel that makes ion emission from grains such a crucial effect.}    

Clearly, the contribution of the dust grains to gas ionization levels depends on the work function $W$ of the grain material. We adopt the fiducial value of \citet{Desch2015}, $W=5$\,eV, and refer the reader to their work for a discussion of the experimental results supporting this choice. This value is close to the ionization potential of potassium, $\rm{IP}=4.34$\,eV, indicating that thermionic emission is important for the production of free electrons in the same temperature range as thermal ionization of potassium. Importantly, for this given value of the work function, grains become negatively charged at high temperatures as a large fraction of potassium evaporating from the grains is in ionized state. This results in a reduction of the effective work function $W_{\rm eff}$, since, for negatively charged grains thermionic emission is higher (eq. \ref{eqn:Weff}).

In this work, we assume that the abundances of hydrogen and potassium atoms are $x_{\rm H}=9.21\times 10^{-1}$ and $x_{\rm K}=9.87\times 10^{-8}$, respectively, relative to the total number density of all atomic particles \citep{Keith2014}, and that the mean molecular weight is $\mu=2.34 m_{\rm H}$\footnote{The total number densities of molecular hydrogen and potassium are then related to the gas density as $n_{\rm H_2} = x_{\rm H}/(2-x_{\rm H}) \rho/\mu$ and $n_{\rm K} = 2 x_{\rm K}/(2-x_{\rm H}) \rho/\mu$, respectively.}. The grain material density is $\rho_{\rm gr}=3.3$\,g\,cm$^{-3}$, the same as in our calculation of the dust opacities. The input for the chemical network are temperature $T$, pressure $P$, hydrogen ionization rate $\zeta$, dust-to-gas ratio $f_{\rm dg}$ and dust grain size $a_{\rm gr}$. All other kinetic rates, parameters and coefficients are the same as in \citet{Desch2015}.\footnote{Note that the fiducial value of the hydrogen ionization rate used by \citet{Desch2015} is in fact $\zeta = 1.4\times10^{-22}$\,s$^{-1}$ (S. Desch, priv. comm.).} We note that the chosen value of the sticking coefficient for electrons ($S_e=0.6$) is compatible with the detailed calculation by \citet{Bai2011b}, \textcolor{red}{who consider work function values} of 1\,eV and 3\,eV. His results suggest that for a work function of 5\,eV, at 1000\,K, $S_e$ is indeed a few times 0.1 for neutral grains, and increases further for negatively-charged grains. For the sticking coefficients for the ions we adopt $S_{j}=1$, \textcolor{red}{and similarly for the neutral atoms we adopt a sticking coefficient of $S_0=1$.}

For a given dust-to-gas ratio $f_{\rm dg}$ and grain size $a_{\rm gr}$, we pre-calculate and tabulate the equilibrium number densities of electrons and ions, and the average grain charge, as functions of temperature $T$, pressure $P$ and hydrogen ionization rate $\zeta$. We find the equilibrium solution following the same method as \citet{Desch2015}. Time derivatives of all number densities are set to zero, and so rate equations yield an algebraic system of equations. For a given average grain charge $Z$, this system of equations is solved iteratively to find number densities of all five species. The grain charge is \textcolor{red}{then} found by solving the equation of charge neutrality.

\subsubsection{Grain size distribution / Effective dust-to-gas ratio} \label{sec:effective_fdg}
The described chemical network incorporates only one grain size population. Ideally, we would consider a number of grain size populations, with the same size distribution used in our calculation of dust opacities. However, this would greatly enhance the computational complexity of the problem. At the same time, it is clear that dust grains of different size contribute differently to the equilibrium ionization levels. To the lowest order of approximation, all dust-related reaction rates are regulated by the total grain surface area. Thus, it can be expected that the ionization levels are most sensitive to the smallest grains. \citet{Bai2009} considered the effects of dust on the ionization levels in the cold regions of protoplanetary discs, in application to the onset of the MRI due to non-thermal sources of ionization. They considered chemical networks with two grain size populations and found that the grain populations behave independently, as the charge transfer between the grains is negligible. They further found that the ionization levels are largely controlled by a quantity $f_{\rm dg}/a_{\rm gr}^p$, where the exponent $p$ varies between $p=1$ (i.e. the total grain surface area) and $p=2$.

\begin{figure*}
    \centering
    \includegraphics[height=0.28\textwidth,trim={0 0 1.5cm 0},clip]{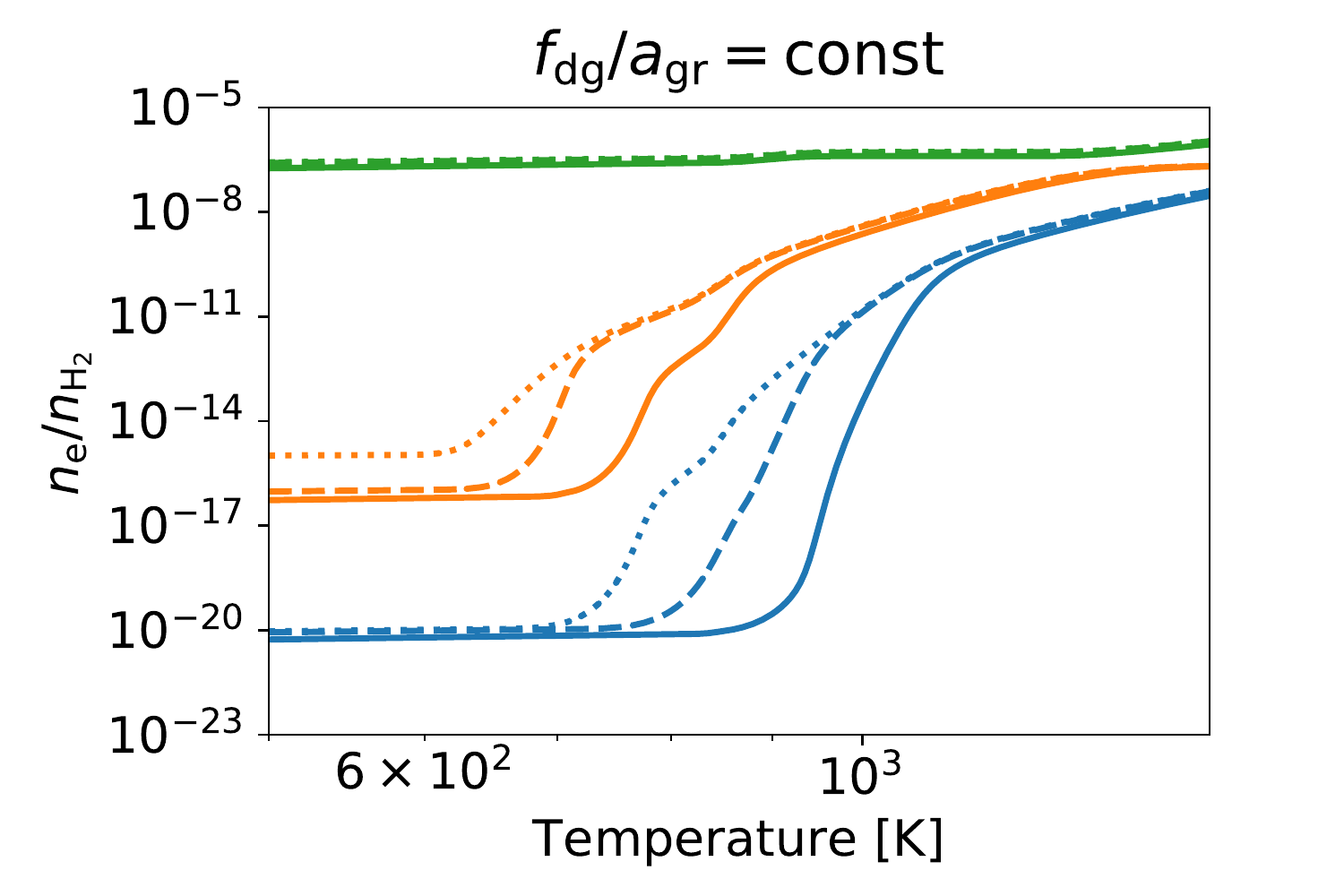}
	\includegraphics[height=0.28\textwidth,trim={3cm 0 1.5cm 0},clip]{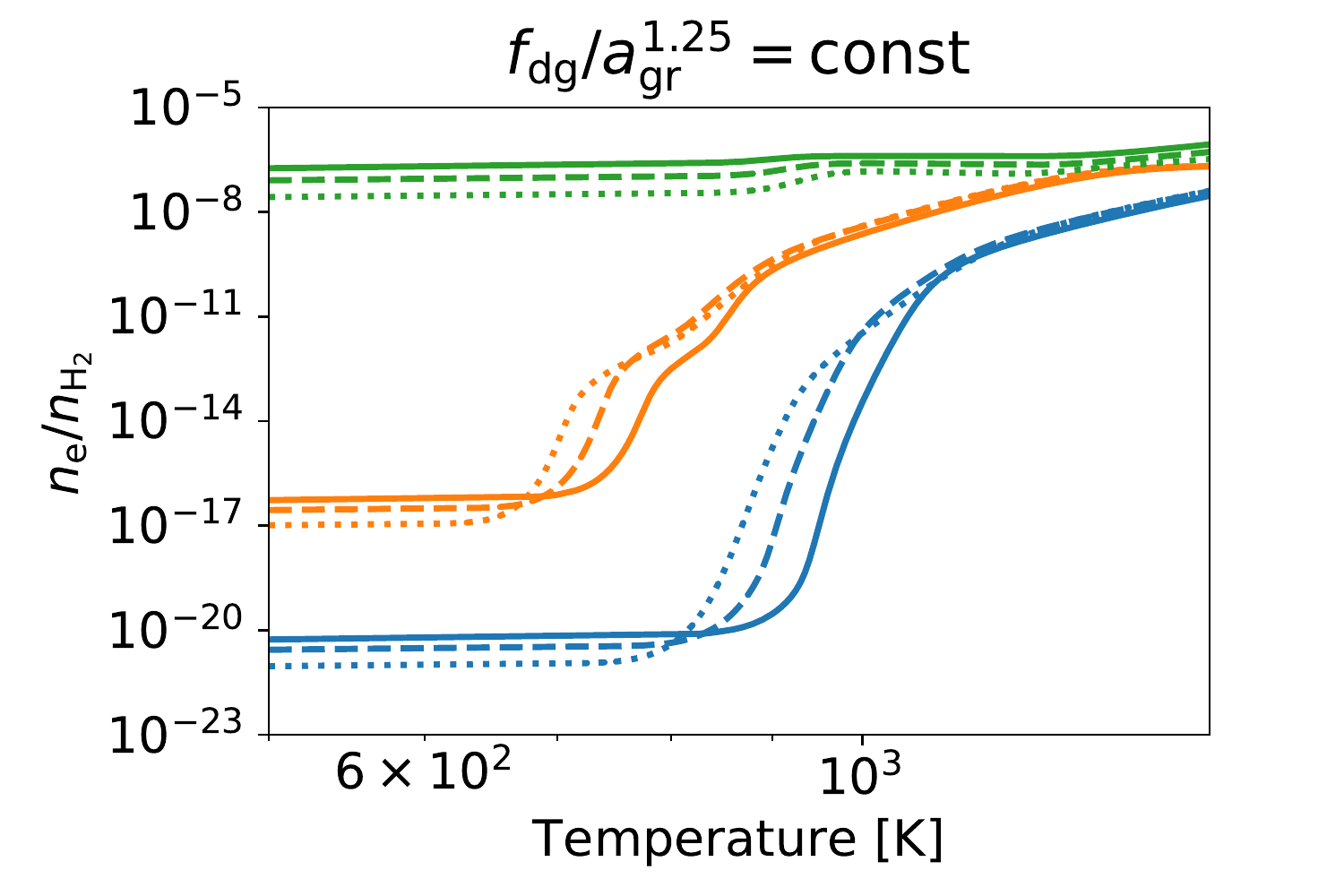}
	\includegraphics[height=0.28\textwidth,trim={3cm 0 1.5cm 0},clip]{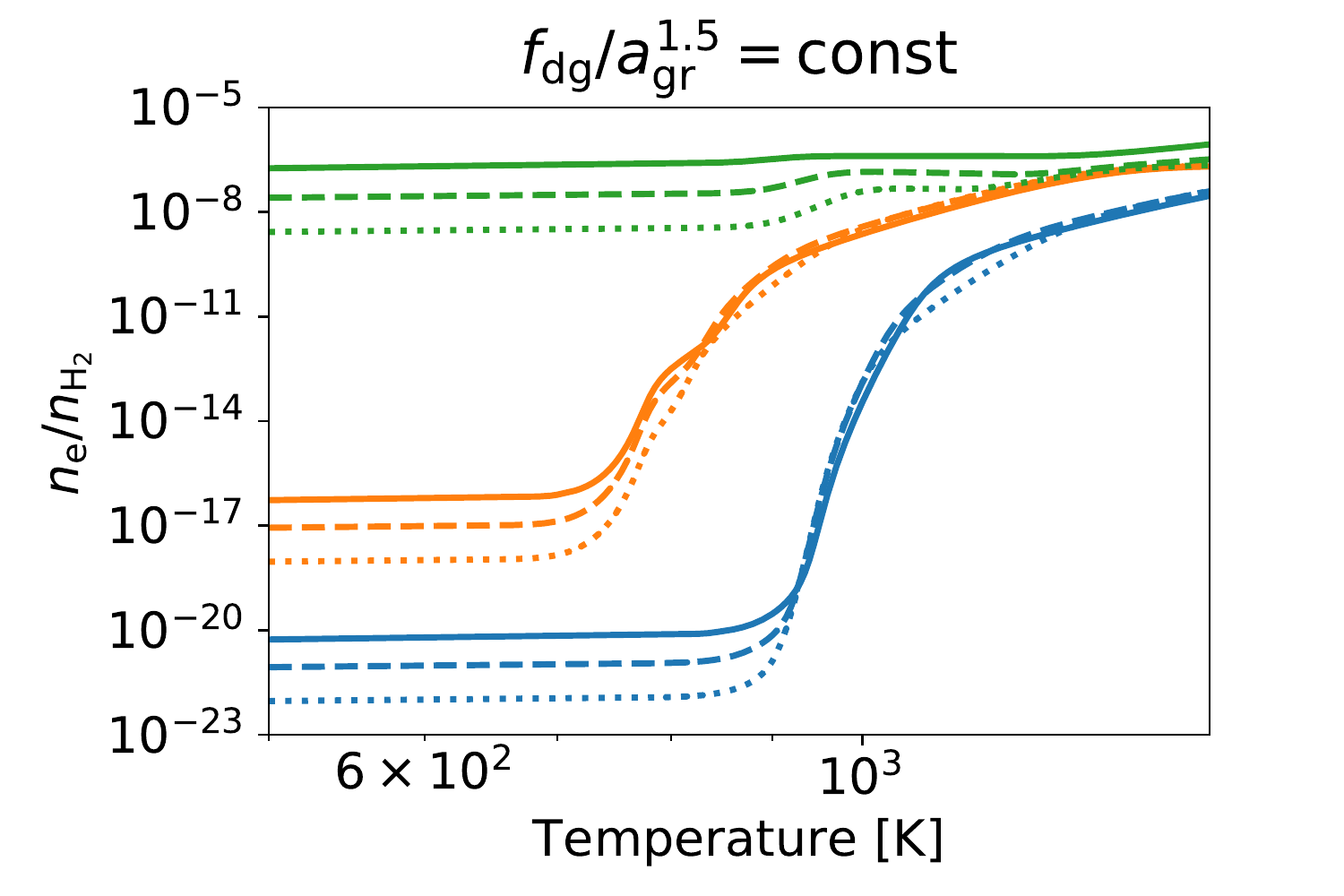}
    \caption{Ionization fraction ($n_{\rm e}/n_{\rm H_2}$) as a function of temperature. Different colors correspond to different combinations of pressure and hydrogen ionization rates: $P=10^2$\,dyn\,cm$^{-2}$ and $\zeta=10^{-19}$\,s$^{-1}$ (blue), $P=10^{-2}$\,dyn\,cm$^{-2}$ and $\zeta=10^{-19}$\,s$^{-1}$ (orange), $P=10^{-2}$\,dyn\,cm$^{-2}$ and $\zeta=10^{-11}$\,s$^{-1}$ (green). Different linestyles correspond to different grain sizes, $a_{\rm gr}=10^{-5}$\,cm (solid), $a_{\rm gr}=10^{-3}$\,cm (dashed), $a_{\rm gr}=10^{-1}$\,cm (dotted), each with a different dust-to-gas ratio such that the ratio $f_{\rm dg}/a_{\rm gr}^p$ remains constant and as evaluated for $a_{\rm gr}=10^{-5}$\,cm, $f_{\rm dg}=0.01$. Different panels show the calculations for different values of the exponent $p$, as indicated in each panel title. Exponent $p=1$ is equivalent to keeping the total grain surface area constant. Adopting exponent $p=1.5$ yields approximately the same threshold temperature at which ionization fraction sharply increases irrespective of grain size $a_{\rm gr}$. See Section \ref{sec:effective_fdg}.}
    \label{fig:ionization}
\end{figure*}

We repeat a similar exercise for the above chemical network, suited to the hot inner regions of protoplanetary discs. For a set of values of the exponent $p$ = 1, 1.25 \textcolor{red}{and} 1.5, we vary the grain size $a_{\rm gr}$ and \textcolor{red}{dust-to-gas mass ratio $f_{\rm dg}$} while keeping $f_{\rm dg}/a_{\rm gr}^p$ constant. We calculate the ionization levels as a function of temperature, and for different sets of pressure and hydrogen ionization rates, so as to probe different conditions in different regions of the inner disc. The results are shown in Fig. \ref{fig:ionization}. The grain surface area ($p = 1$) controls the equilibrium ionization levels when the hydrogen ionization dominates, but does not determine the temperature at which the ionization levels rise due to thermionic and ion emission. On the other hand, for $p=1.5$, this temperature depends very weakly on the grain size. That is, regardless of the actual dust grain size, a quantity $f_{\rm dg}/a_{\rm gr}^{1.5}$ regulates the temperature at which dense interior of the disc becomes ionized. 

Therefore, \textcolor{red}{for the chemical network calculations}, we use a single grain size of $a_{\rm gr}=10^{-5}$\,cm, \textcolor{red}{but employ an effective dust-to-gas ratio $f_{\rm dg, eff}$ that satisfies} $f_{\rm dg, eff} a_{\rm gr}^{-p} = \int_{a_{\rm min}}^{a_{\rm max}} dn(a) m(a)/\rho_{\rm g} a^{-p}$, where $n(a)$ is the same grain size distribution used to calculate dust opacities. Since we find that thermionic and ion emission are more important than \textcolor{red}{the non-thermal ionization of hydrogen} in the inner disc, we use $p=1.5$. From \textcolor{red}{the right panel of} Fig. \ref{fig:ionization}, it appears that this choice will lead to a large error \textcolor{red}{in the gas ionization state we derive} in the low-density non-thermally ionized disc regions, \textcolor{red}{due to the large variation in such regions in the ionization state produced by different grain sizes . However, in reality the error will be much smaller than implied by Fig. \ref{fig:ionization}, since the majority of grains are skewed towards small sizes in a realistic grain size distribution.} 

\subsubsection{Hydrogen ionization rate}
\label{Hionizationrate}
In our calculation of the MRI-driven viscosity, we consider molecular hydrogen ionization rate due to radionuclides, cosmic rays and stellar X-rays. The ionization rate of molecular hydrogen due to short-lived and long-lived radionuclides is 
\begin{equation} 
    \zeta_{\rm R} = 7.6 \times 10^{-19}\textrm{\,s}^{-1},
\end{equation}
which predominantly comes from decay of $^{26}$Al \citep{Umebayashi2009}. The ionization rate of molecular hydrogen due to interstellar cosmic rays is \citep{Umebayashi2009}
\begin{equation} 
    \zeta_{\rm CR}(z) = \frac{\zeta_{\rm CR,ISM}}{2} e^{-\frac{\Sigma(z)}{\lambda_{\rm CR}}} \left( 1 + \left( \frac{\Sigma(z)}{\lambda_{\rm CR}} \right)^{\frac34} \right)^{-\frac43},
\end{equation}
where $\zeta_{\rm CR,ISM} = 10^{-17}$\,s$^{-1}$ is the interstellar cosmic ray ionization rate, $\Sigma(z)$ is the integrated density column from the top of the disc to the height $z$ above disc midplane, and $\lambda_{\rm CR} = 96$\,g\,cm$^{-2}$ is the attenuation length for cosmic rays \citep{Umebayashi1981}.

For the ionization rate of molecular hydrogen due to stellar X-rays we use \citet{Bai2009} fits to the \citet{Igea1999} Monte Carlo simulations,
\begin{align} 
\zeta_{\rm X}(z) = \frac{L_{\rm X}}{10^{29}\textrm{\,erg\,s}^{-1}} \left( \frac{r}{1\textrm{\,AU}} \right)^{-2.2} \large( \zeta_1 e^{-(\Sigma(z)/\lambda_1)^{c_1}} + \\ \nonumber
\zeta_2 e^{-(\Sigma(z)/\lambda_2)^{c_2}} \large),
\end{align}
where $L_{\rm X}$ is stellar X-ray luminosity, $\zeta_1 = 6 \times 10^{-12}$\,s$^{-1}$, $\lambda_1 = 3.4 \times 10^{-3}$\,g\,cm$^{-2}$, and $c_1 = 0.4$ characterise absorption of X-rays, and $\zeta_2 = 10^{-15}$\,s$^{-1}$, $\lambda_2 = 1.6$\,g\,cm$^{-2}$, and $c_2 = 0.65$ characterise the contribution from scattered X-rays. Here we have re-calculated the attenuation lengths given by Bai \& Goodman in terms of column densities of hydrogen nucleus into the surface density lengths using the hydrogen abundance given above. We adopt the saturated relationship $L_{\rm X} = 10^{-3.5} L_{\rm bol}$ \citep[e.g.][]{Wright2011}. For both cosmic rays and X-rays, we ignore the contribution coming through the other side of the disc. This is valid since we find that the gas surface densities are mostly larger than the attenuation lengths of the ionizing particles and, even at low gas surface densities, this can only increase the ionization rates by at most a factor of 2.

\subsection{Numerical methods} \label{sec:numerical_methods}
\subsubsection{Equilibrium vertical disc structure}
At a given orbital radius, magnetic field strength and grazing angle $\phi$, the disc's vertical structure is determined as a solution to the boundary value problem given by eqns. (\ref{eq:hydrostatic_equilibrium}), (\ref{eq:rt_1})-(\ref{eq:rt_3}) and the ideal gas law, and, where convectively unstable, by eqns. (\ref{eq:hydrostatic_equilibrium}), (\ref{eq:rt_1}) and (\ref{eq:convection}). This boundary value problem is solved using the shooting method \citep{Press2002}. The equations are integrated from the top of the disc ($z=z_{\rm surf}$) to the disc midplane ($z=0$). The height of the disc $z_{\rm surf}$ is then found, such that $F(0)=0$. This root-finding problem is solved using the Ridders' method, with an exit criterion that $|F(0)|<10^{-5}F(z_{\rm surf})$.

Equations of the vertical disc structure are integrated on a fixed vertical grid (i.e., in the $z$-direction), with points uniform in the polar angle. Since eq. (\ref{eq:rt_3}) is an algebraic equation, and for numerical stability, we use a fully implicit integration method. For the radiative-transfer problem we use a Runge-Kutta method of the second order, i.e. the trapezoidal method. In the trapezoidal method the equations (\ref{eq:hydrostatic_equilibrium}), (\ref{eq:rt_1})-(\ref{eq:rt_3}) are discretised as a system of nonlinear equations to be solved in every integration step (a system of nonlinear equations in $P_{n+1}$, $F_{n+1}$, $J_{n+1}$, $T_{n+1}$ to be solved by a root-finding algorithm),
\begin{align} \nonumber
	P_{n+1} &= P_n +
	\frac{h}{2} \Omega^2 ( - \rho_n z_n - \rho_{n+1} z_{n+1} ) , \\
	\nonumber
	F_{n+1} &= F_n + \frac{h}{2} (\Gamma_n + \Gamma_{n+1}) , \\
	\nonumber
	J_{n+1} &= J_n + \frac{h}{2} (-\frac{3}{4\pi}) (\rho_n \kappa_{\rm R}(T_n) F_n + \rho_{n+1} \kappa_{\rm R}(T_{n+1}) F_{n+1}) , \\
	\nonumber
	0 &= 4 \rho_{n+1} \kappa_P(T_{n+1}) (\sigma_{\rm SB} T_{n+1}^4 - \pi J_{n+1}) - \Gamma_{n+1} ,
\end{align}
where $h$ is the integration step, $\rho=\rho(T, P)$ is given by the ideal gas law and $\Gamma = \Gamma_{\rm acc} + \Gamma_{\rm irr}$. Here, $\Gamma_{{\rm irr},n}=\Gamma_{{\rm irr},n}(T_n, P_n, \tau_{\textrm{irr},z,n})$, with the optical depth to stellar irradiation obtained using $\tau_{\textrm{irr},z,n+1}=\tau_{\textrm{irr},z,n} + \frac{h}{2} \kappa_{\rm P}^* (\rho_n + \rho_{n+1})$. Furthermore, viscous dissipation is a function of the MRI-driven viscosity $\alpha$ and thus depends on the local ionization levels. The latter are a function of the local temperature, pressure and the hydrogen ionization rate which depends on the column density from the top of the disc. Thus, $\Gamma_{{\rm acc},n}=\Gamma_{{\rm acc},n}(T_n, P_n, N_n)$, with the column density given by $N_{n+1} = N_n + \frac{h}{2} k_{\rm B}\textcolor{red}{^{-1}} (P_n/T_n + P_{n+1}/T_{n+1})$.

The equation for pressure $P_{n+1}$ can be rearranged into an explicit form
\begin{align} \nonumber
	P_{n+1} &= \frac{ P_n - \frac{h}{2} \Omega^2 \rho_n z_n }{ 1 + \frac{h}{2} \Omega^2 \frac{\mu m_{\rm H}}{k_{\rm B}} \frac{z_{n+1}}{T_{n+1}} }.
\end{align}
Then, the above system of equations is equivalent to a single nonlinear equation in $T_{n+1}$, greatly simplifying the problem. In every integration step we use the Ridders' method to solve this equation for the temperature $T_{n+1}$ (down to a relative precision of $10^{-7}$) and consequently for all other quantities. This includes the MRI-driven viscosity $\alpha$, which is thus calculated self-consistently at each step of integration.\footnote{This is indeed necessary. An iterative method in which the disc thermal structure is decoupled from the density structure and the heating terms (e.g. Dullemond et al. 2002) does not converge to a solution in the case of MRI-driven viscosity.} At every step and in every iteration of the root-solver opacities are interpolated from pre-calculated tables using cubic splines, and the ionization levels (e.g., free electron number density) using tri-linear interpolation.


Additionally, at each integration step we check if the resulting temperature gradient is unstable to convection, and if so, the temperature $T_{n+1}$ is obtained analytically using
\begin{align} \nonumber
	T_{n+1} &= T_n + \frac{h}{2} \nabla_{\rm ad} \frac{\mu m_{\rm H}}{k_{\rm B}} \Omega^2 (- z_n - z_{n+1}) .
\end{align}
With $T_{n+1}$ known, all other quantities follow same as above. A disc column can, in principle, become convectively stable again at some height above disc midplane. To calculate the mean intensity $J$ at a boundary between a convective and a radiative zone, we use the energy balance equation (\ref{eq:rt_3}).

For some model parameters there can be a range of orbital radii and values of the magnetic field strength for which there are multiple solutions for the disc height $z_{\rm surf}$ (i.e. multiple solutions for the equilibrium vertical disc structure). This happens when a complex ionization structure leads to strong variations in the viscosity $\alpha$ as a function of height above disc midplane, making the total produced viscous dissipation a non-monotonous function of $z_{\rm surf}$. When there are multiple solutions, we choose the solution with smallest $z_{\rm surf}$. It is likely that at least some of the additional solutions are thermally unstable and/or unphysical, as the strong variations in both the levels of turbulence and the levels of ionization should be removed by turbulent mixing (see section \ref{sec:multiple_solutions}). \textcolor{red}{Note that the multiple solutions in our model, when they exist, correspond to the same, radially-constant input accretion rate. Steady accretion is thus assured regardless of the choice of the solution.}

\subsubsection{Magnetic field strength} \label{sec:magnetic_field}
At a given orbital radius, for a fixed grazing angle, $\phi$, and magnetic field strength, $B$, the above procedure yields an equilibrium vertical disc structure characterised by a vertically-averaged viscosity
\begin{equation} \nonumber
    \bar\alpha = \frac{ \int_0^{z_{\rm surf}} \alpha P dz }{ \int_0^{z_{\rm surf}} P dz }.
\end{equation}
As in the vertically-isothermal model \citep{Mohanty2018}, for a sufficiently small and a sufficiently large magnetic field strength $B$ the MRI is suppressed in the entire disc column and $\bar\alpha=\alpha_{\rm DZ}$. There can be an intermediate range of magnetic fields strengths for which the MRI is active, and the vertically-averaged viscosity $\bar\alpha$ peaks at some value of $B$. At every orbital radius, we choose $B$ such that $\bar\alpha$ is maximized. \textcolor{red}{The underlying assumption here is that the magnetic fields are strengthened by the MRI-driven turbulence. It is assumed that the initial magnetic field configuration was sufficient to start the instability. The induced magnetohydrodynamic turbulence then amplifies the field strength. We assume the equilibrium configuration is one in which the turbulence is maximized, where the turbulence is parametrized by $\bar\alpha$. Similar arguments were employed by \citet{Bai2011b} and \citet{Mohanty2013}.} 

\textcolor{red}{To maximize $\bar\alpha(B)$, we use the Brent method with a target absolute precision of $10^{-3}$ in $\textrm{log} B$. There can be a range of orbital radii where there are multiple local maxima in $\bar\alpha$ as a function of $B$.} This is essentially for the same reasons that cause multiple solutions in disc height $z_{\rm surf}$ at a fixed value of $B$. In general we choose $B$ corresponding to the global maximum in $\bar\alpha$. However, in some cases, this is a function of the grazing angle $\phi$ at a fixed orbital radius, and the procedure to determine the grazing angle (described below) does not converge. There we choose a local maximum with a largest magnetic field strength.

\subsubsection{Grazing angle}
At any given orbital radius, the angle between the incident stellar radiation and the disc surface is determined self-consistently with the disc structure following \citet{DAlessio1999}. A self-consistent disc structure is found by iteratively updating the grazing angle and re-calculating the entire disc structure. We use a logarithmic grid for orbital radius. First, we calculate $\phi$ using eq. (\ref{eq:graz_angle}) by assuming that $z_{\rm irr}=0$ and solve for the vertical disc structure and the magnetic field strength at all radii. We integrate through the obtained disc structure along lines-of-sight to the star to calculate $\tau_{\rm irr}(r, z)$, which yields an updated $z_{\rm irr}$ at each radius.

Critically, to calculate the updated value of the grazing angle $\phi$ at each radius, the derivative $d \textrm{log} z_{\rm irr}/d \textrm{log} r$ is approximated by assuming that $z_{\rm irr}$ is a power-law, $z_{\rm irr}$\,$\propto$\,$r^b$, within a radius bin centered at the given radius. So, at each radius we fit for the slope $b$ using $z_{\rm irr}$ at that radius and at a number of radial grid points interior and exterior to it. Then, the value of the grazing angle is updated and the vertical disc structure re-calculated at all radii.

This procedure is repeated until the grazing angle has converged at every radius, i.e., until the relative difference in $\phi$ between two consecutive iterations is less than 0.5\% at all radii. For the first radial point we always assume the flat-disc approximation ($z_{\rm irr}=0$) and do not include it in the fitting routine. In this work we use a total of 100 radial points between 0.1\,AU and 1\,AU, and a total of 10 radial points in fits for $d \textrm{log} z_{\rm irr}/{d \textrm{log} r}$.


\section{Results} \label{sec:results}

Using our self-consistent model we can now investigate the structure of the inner regions of a protoplanetary disc that is viscously accreting with a constant accretion rate. 

In section \ref{sec:thermal_struct} we consider a model in which the vertical structure of the disc is calculated self-consistently from viscous heating, heating by stellar irradiation, radiative and convective energy transport, and self-consistent radiative properties of dust, while the disc's ionization state is calculated by only considering the thermal (collisional) ionization of potassium, using the Saha equation. 

In section \ref{sec:chemical_struct} we present results for our full model that includes \textcolor{red}{additional chemical species, including dust grains, in the chemical network,} as well as non-thermal sources of ionization.

Unless otherwise stated, throughout this paper we assume a solar-mass star, $M_*=1$\,M$_\odot$, with a stellar radius $R_*=3$\,R$_\odot$, effective stellar temperature $T_*=4400$\,K, gas accretion rate $\dot{M}=10^{-8}$\,M$_\odot$\,yr$^{-1}$, and viscosity in the MRI-dead zone $\alpha_{\rm DZ}=10^{-4}$. Our adopted gas accretion rate is the median from observations of solar-mass classical T Tauri stars \citep[e.g.,][]{Hartmann1998,Manara2016,Manara2017}. \textcolor{red}{In reality, there is more than an order of magnitude scatter in the data around this median; we explore the effects of varying $\dot{M}$, as well as other parameters, in a companion paper, as noted below.}  The stellar parameters are from the evolutionary models of \citet{Baraffe2015}, for a solar-mass star at an age of $5\times10^5$\,yr. At later times, the stellar luminosity decreases, as the star contracts towards the zero-age main-sequence. 
Our choice of a relatively young classical T Tauri star thus maximises the stellar luminosity, allowing us to examine the largest possible effect of stellar irradiation on the inner disc. Lastly, we adopt a standard ISM dust-to-gas ratio of $f_{\rm dg}=10^{-2}$, and a maximum dust grain size of $a_{\rm max}=1$\,$\mu$m. 

\textcolor{red}{The above set of parameters represents our {\it fiducial model}. In this work we investigate in detail the main physical and chemical processes that shape the structure of the inner disc for this model. In a companion paper (Jankovic et al. {\it in prep.}), we examine how the inner disc structure changes as a function of the model parameters, and build a picture of how this structure may affect planet formation.} 

\subsection{Disc thermal structure and the MRI} \label{sec:thermal_struct}
In this section we explore how thermal processes shape the inner disc structure. The ionization state of the disc is set here only by the thermal ionization and recombination of potassium, which is assumed to be entirely in the gas phase. Note that there is no other chemistry in this model: in the absence of grain-related reactions (grains here only contribute to the opacity), and without non-thermal sources to ionize H$_2$, gas-phase thermal ionization and recombination of potassium are the only reaction pathways available. 

\citet{Mohanty2018} presented a model in which the disc is vertically isothermal, with a constant opacity (=10\,cm$^2$\,g$^{-1}$). Throughout this section, we shall compare this simple model with three models of increasing complexity within our framework: {\it (i)} a model with the same constant opacity ($\kappa_{\rm R}=\kappa_{\rm P}=10$\,cm$^2$\,g$^{-1}$), but with the disc vertical structure calculated self-consistently from viscous heating and cooling by radiation and convection; {\it (ii)} a model with the same heating and cooling processes as (i), but now with opacities also determined self-consistently; and {\it (iii)} similar to (ii), but now also including heating by stellar irradiation. 

\textcolor{red}{To make it easier to interpret the effects of these thermal processes, we first focus on how the disc's temperature changes when the increasingly complex models are used. We then discuss how this leads to differences in the disc's ionization state, \textcolor{red}{the locations where} the MRI is active, and the radial surface density and midplane pressure profiles. \textcolor{red}{Finally we} provide a short summary of the key findings at the end of this section.}

\subsubsection{Thermal structure of the inner disc} \label{sec:thermal_struct_disc}
\begin{figure*}
    \centering
    \includegraphics[height=0.26\textwidth,trim={0 0 3.3cm 0},clip]{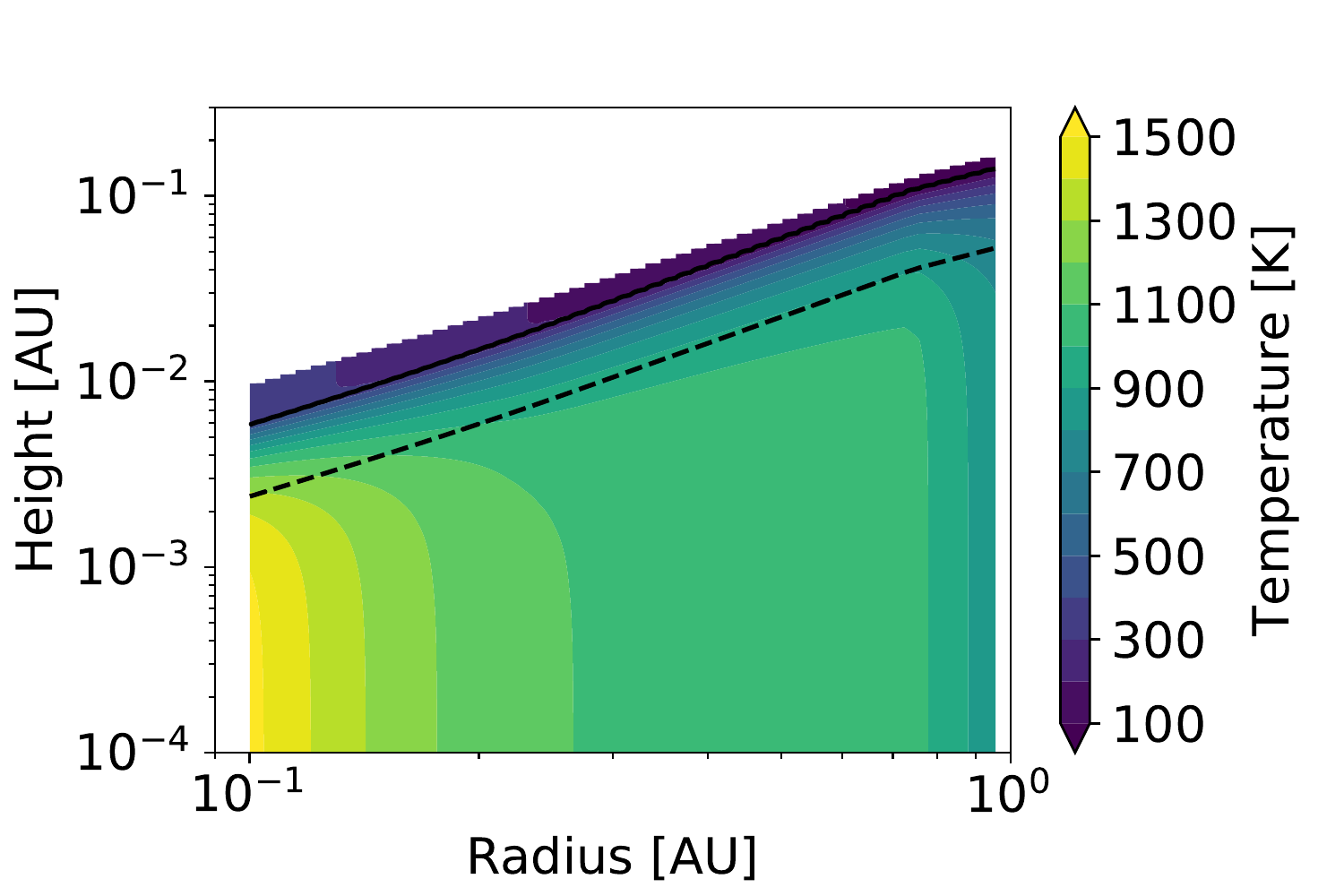}
	\includegraphics[height=0.26\textwidth,trim={2.2cm 0 3.3cm 0},clip]{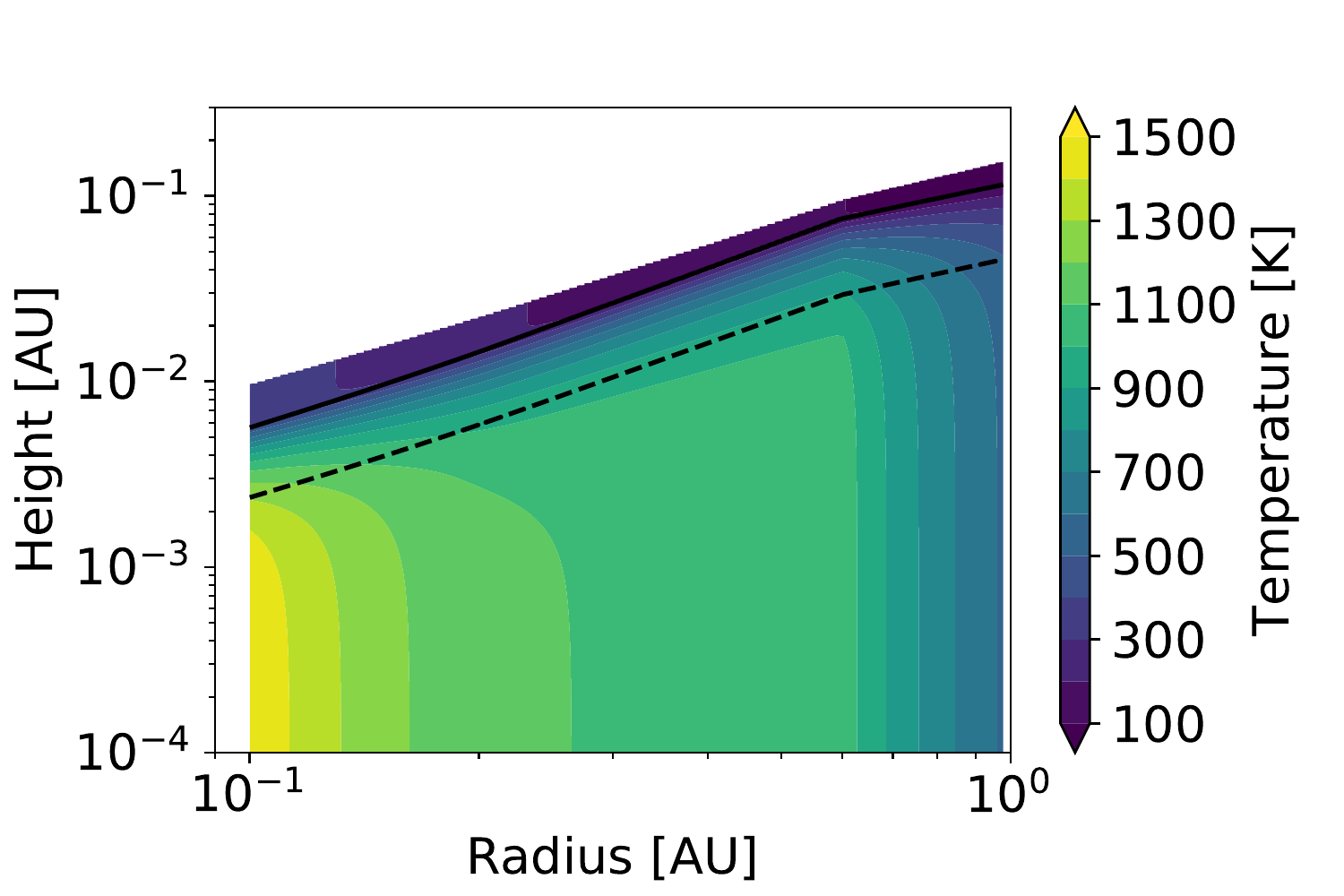}
	\includegraphics[height=0.26\textwidth,trim={2.2cm 0 0 0},clip]{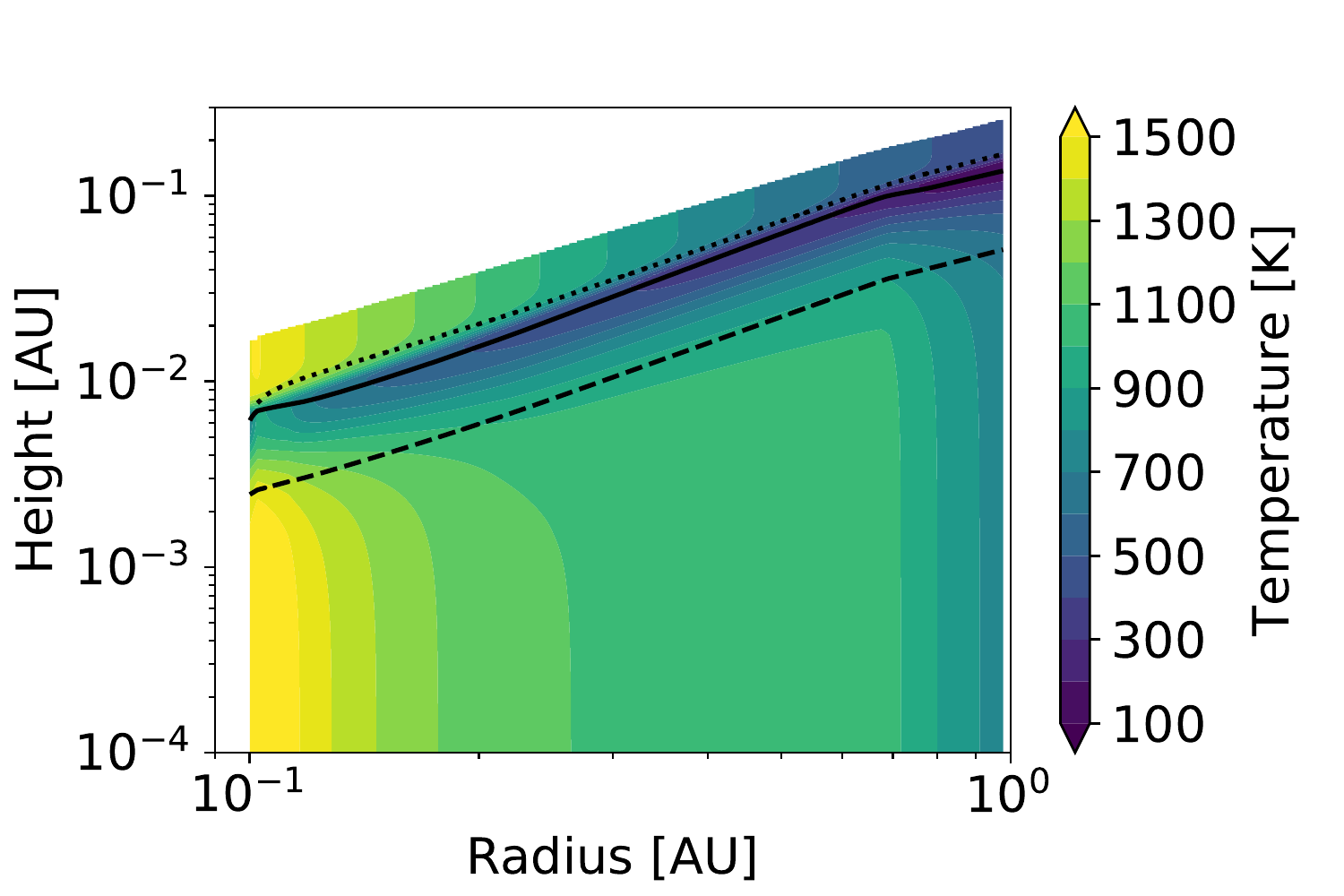}
    \caption{Temperature as a function of location in the disc for a viscously-heated constant-opacity model (left), viscously-heated model with realistic opacities (middle), and a viscously and irradiation-heated model with realistic opacities (right). In each panel the solid line shows the disc photosphere ($\tau_{\rm R}=2/3$) and the dashed line shows the pressure scale height ($P=e^{-1/2}P_{\rm mid}$). The dotted line in the right-hand-side panel shows the surface at which $\tau_{\rm irr}=2/3$. Note that the inclusion of heating by stellar irradiation does not strongly affect the disc midplane temperature. See Section \ref{sec:thermal_struct_disc}.}
    \label{fig:thermal_temperature}
\end{figure*}

Figure \ref{fig:thermal_temperature} shows the temperature as a function of disc radius and height above the midplane for our three models, increasing in complexity from the left to right panels.

Noticeably, the temperature profiles deviate from being vertically isothermal. The dashed lines here show the disc pressure scale height, and the solid lines show the disc photosphere for outgoing radiation (i.e., where the Rosseland-mean optical depth is $\tau_{\rm R}=2/3$). In all three models the temperature increases towards the midplane below the photosphere, as the disc becomes more and more optically-thick to its own radiation.

\begin{figure*}
    \centering
    \includegraphics[height=0.26\textwidth,trim={0 0 1.0cm 0},clip]{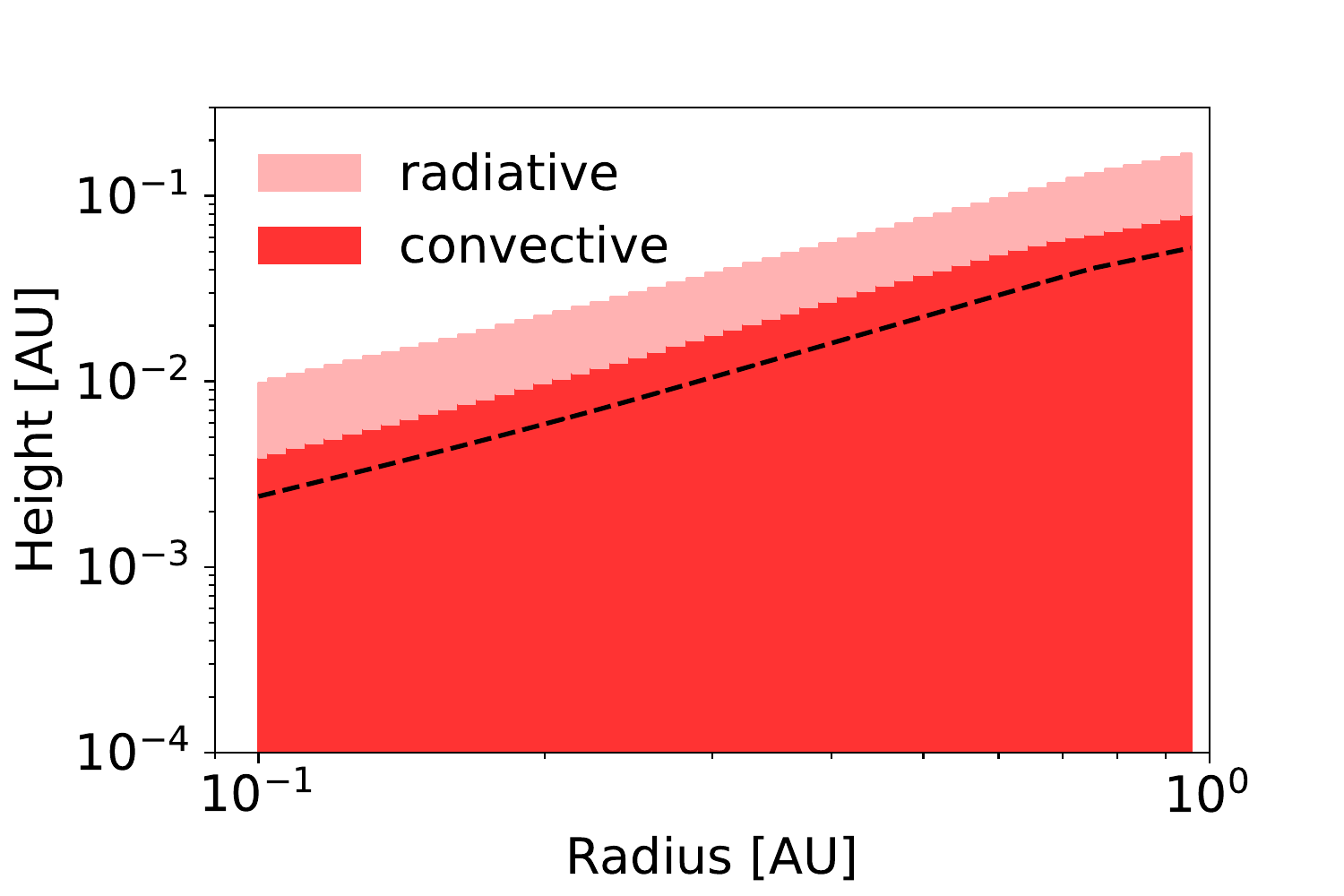}
	\includegraphics[height=0.26\textwidth,trim={2.2cm 0 1.0cm 0},clip]{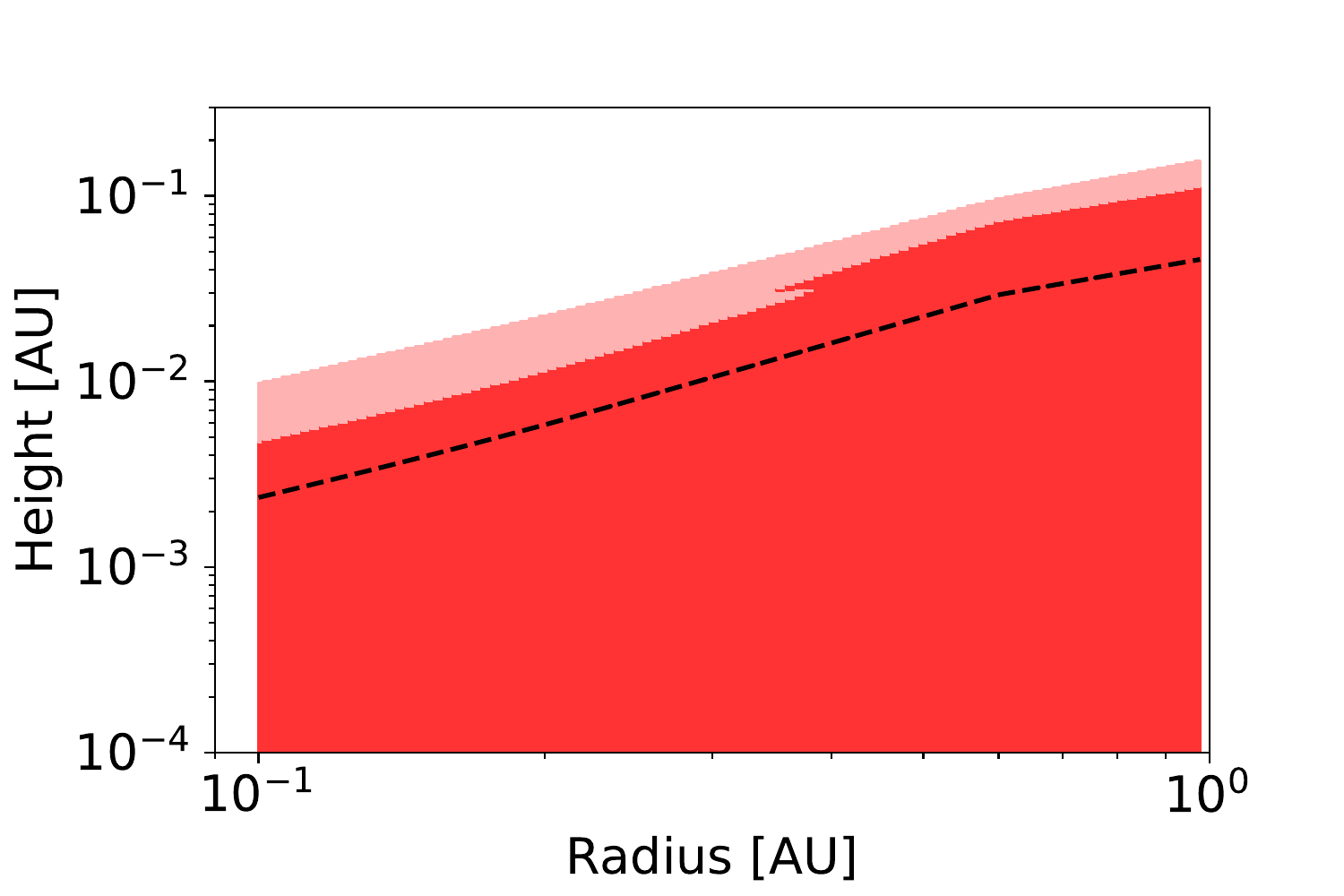}
	\includegraphics[height=0.26\textwidth,trim={2.2cm 0 1.0cm 0},clip]{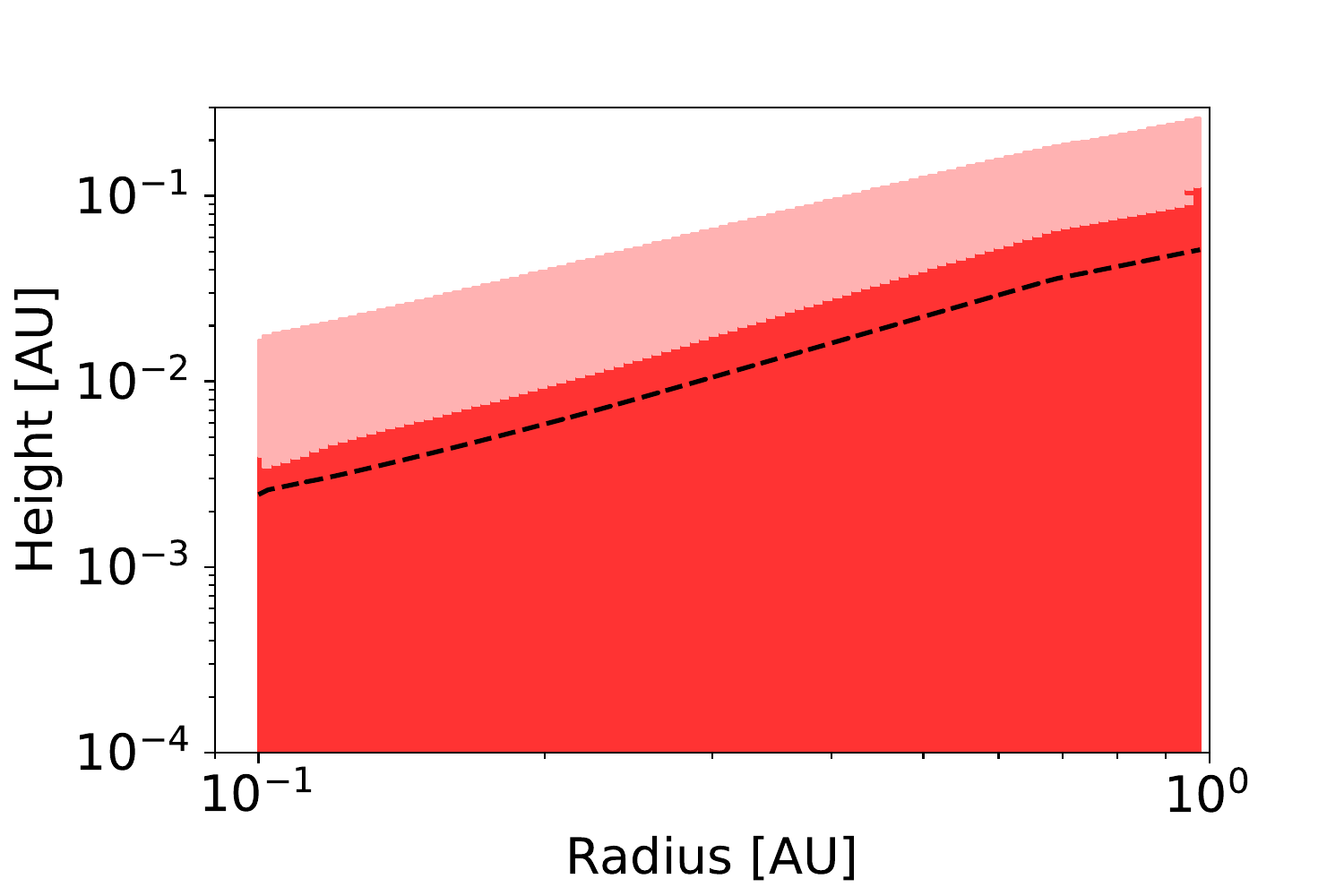}
    \caption{Radiative and convective zones (light and dark red, respectively) for a viscously-heated constant-opacity model (left), viscously-heated model with realistic opacities (middle), and a viscously and irradiation-heated model with realistic opacities (right). In each panel the dashed line shows the pressure scale height ($P=e^{-1/2}P_{\rm mid}$). In all three models the disc is convectively unstable within few scale heights. See Section \ref{sec:thermal_struct_disc}.}
    \label{fig:thermal_convection}
\end{figure*}

The resulting temperature gradient is sufficiently high to make the disc convectively unstable, as shown in Fig. \ref{fig:thermal_convection}. From the midplane up to a couple of pressure scale heights, energy is thus transported by convection, and the temperature gradient here is essentially isentropic. Importantly, the strong temperature gradient that yields this convective instability is not specific to MRI-driven accretion: it is a {\it general} feature of any active, optically thick disc where the accretion heat is released near the midplane \citep[see also][]{Garaud2007}. We show this analytically in section \ref{sec:discussion}.

The model that includes heating by stellar irradiation, shown in the right panels of Figs. \ref{fig:thermal_temperature} and \ref{fig:thermal_convection}, features a temperature inversion in the disc upper layers. This inversion has been discussed in detail by \citet{DAlessio1998}, and is a simple consequence of the fact that the photosphere for the outgoing radiation (solid line) lies below the photosphere for the incoming stellar radiation (dotted line), where the latter corresponds to the irradiation surface $z_{\rm irr}$ at which $\tau_{\rm irr}=2/3$. Above this surface the disc upper layers are heated by stellar irradiation. Below, the disc becomes optically thick to stellar radiation (i.e., stellar photons do not penetrate here) and the temperature drops. Going deeper still, below the disc photosphere to the outgoing radiation, the disc becomes optically thick to its own radiation and the temperature rises again.

\begin{figure}
    \centering
    \includegraphics[width=0.5\textwidth, trim={0 2.2cm 0 2.2cm}, clip]{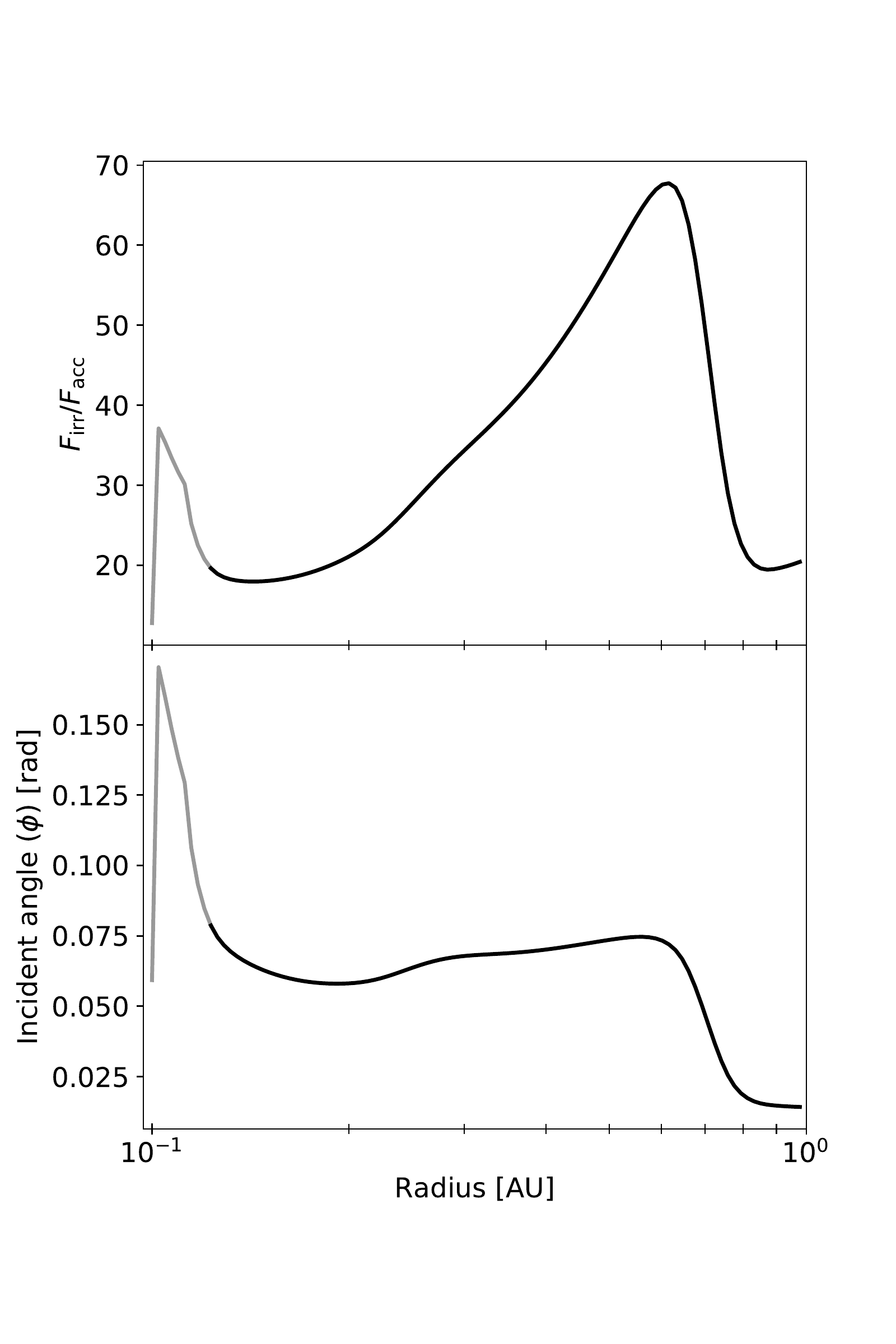}
    \caption{{\it Top:} Ratio of the total irradiation heating to the total viscous dissipation ($F_{\rm irr}/F_{\rm acc}$) as a function of radius. The total (vertically-integrated) absorbed stellar flux $F_{\rm irr}$ is at least an order of magnitude larger than the total viscous dissipation $F_{\rm acc}$ at any given radius. \textcolor{red}{{\it Bottom:} Grazing angle of stellar irradiation} ($\phi$) as a function of radius. \textcolor{red}{Note that the $F_{\rm irr}/F_{\rm acc}$ curve follows the trend in $\phi$. In both panels, the  gray portion of the curve} indicates the region affected by the inner boundary condition on $\phi$. See Section \ref{sec:thermal_struct_disc}.}
    \label{fig:thermal_irradiation}
\end{figure}

In the irradiated disc, close to the star the disc midplane is as hot as the disc upper layers, and further away the midplane is significantly hotter than the upper layers. It would thus appear that accretion heating dominates in the inner disc. However, at a given radius, the total flux of stellar radiation absorbed by a vertical column in the disc is in fact \textcolor{red}{at least an order of magnitude greater than} the total flux generated by viscous dissipation ($F_{\rm irr}$\,$\gtrsim$\,$10 F_{\rm acc}$), throughout the inner disc \textcolor{red}{(see top panel of Fig. \ref{fig:thermal_irradiation})}. In spite of this, the disc temperature near the midplane is only weakly affected by irradiation. We discuss this result in section \ref{sec:importance_irradiation}.

The ratio $F_{\rm irr}/F_{\rm acc}$ varies non-monotonically, following the grazing angle $\phi$, shown in the bottom panel of Fig. \ref{fig:thermal_irradiation}. At the inner edge of our calculation domain, the first $\sim$10 points (shown in gray) are affected by boundary effects. This is a well-known problem in disc models that account for stellar irradiation using the grazing angle prescription \citep{Chiang2001}. Importantly, far from the inner edge, the value of the grazing angle and the disc structure do not depend on the disc structure at the inner edge. 

\subsubsection{Ionization levels and non-ideal MHD effects} \label{sec:ion_levels_non_ideal}
\begin{figure*}
    \centering
    \includegraphics[width=\textwidth,trim={0 0 0 0}, clip]{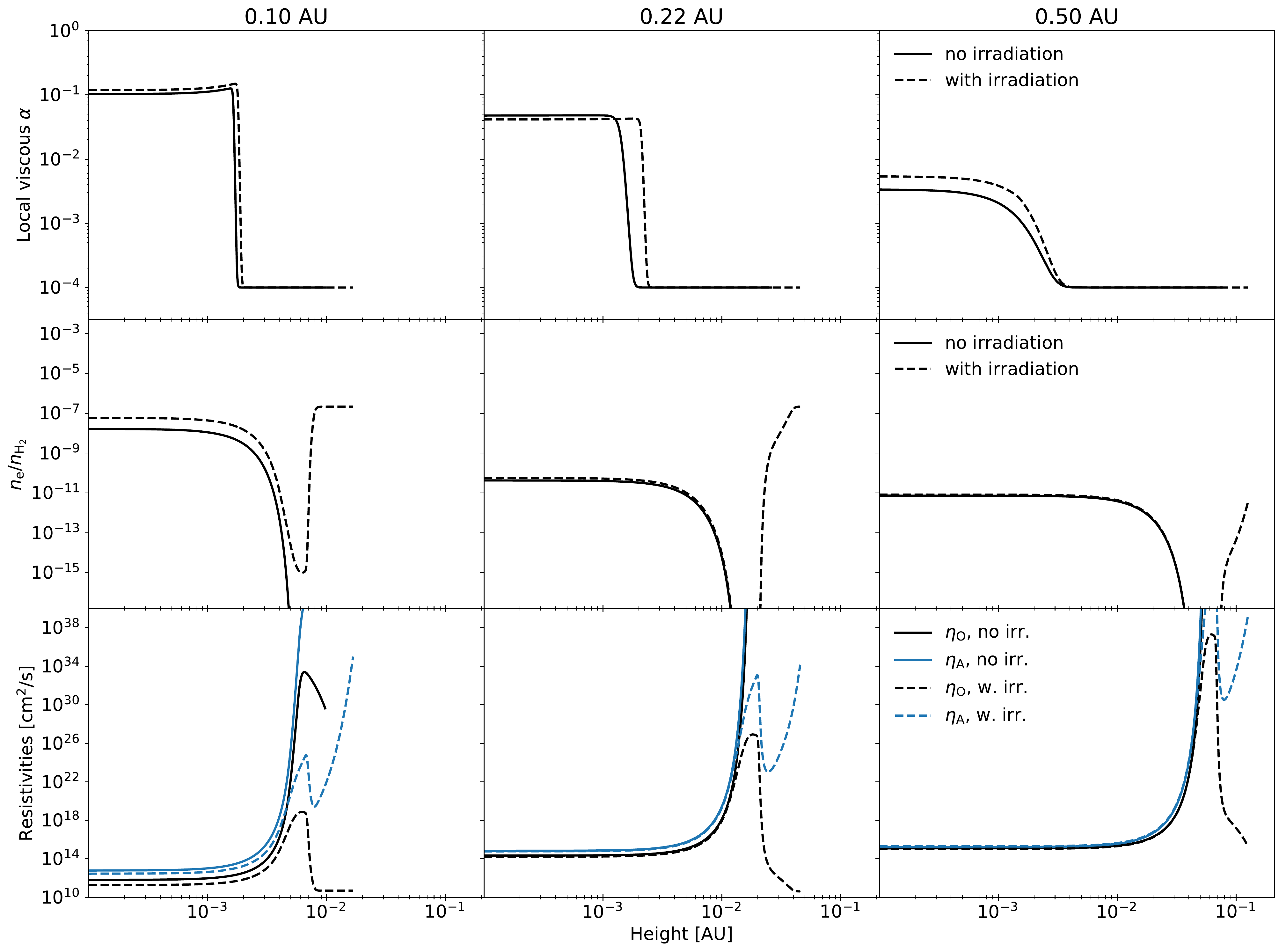}
    \caption{Local viscous parameter $\alpha$ (top), ionization fraction $n_{\rm e}/n_{\rm H_2}$ (middle) and magnetic resistivities (bottom) at three different disc radii (as indicated in the panel titles), \textcolor{red}{in two different models: one with realistic opacities but no irradiation (solid lines), and one that also includes irradiation (dashed lines)}. Despite the high ionization fraction in the irradiation-heated disc upper layers, \textcolor{red}{$\eta_A$ is still very high in these layers, i.e.,} ambipolar diffusion quenches the MRI there. See Section \ref{sec:ion_levels_non_ideal}.}
    \label{fig:thermal_ionization}
\end{figure*}

The ionization structure in our models with a self-consistent vertical structure is qualitatively different from that in the vertically isothermal model of \citet{Mohanty2018}. In the latter, the temperature is constant with height above the midplane while the density decreases, and thus the ionization fraction ($n_{\rm e}/n_{\rm H_2}$) increases with height (by the Saha equation). In our models without stellar irradiation heating, the temperature declines monotonically with height, and hence the ionization fraction decreases as well (Fig. \ref{fig:thermal_ionization}: middle row, solid curves). When we include irradiation heating, on the other hand, the fractional ionization does increase in the uppermost irradiated layers, due to the rising temperature there (dashed curves).

As a consequence of the above, both ambipolar and Ohmic resistivities increase with height in our non-irradiated discs. In other words, in these models, both ambipolar and Ohmic diffusion quench the MRI from above. In the irradiated disc, the Ohmic resistivity falls in the uppermost layers; however, ambipolar diffusion still increases in these layers, again stifling the MRI there. Consequently, the vertical extent of the MRI-active region (i.e., where the local $\alpha$ > $\alpha_{\rm DZ}$) at any given radius in our irradiated disc model is nearly identical to that in the non-irradiated discs (Fig. \ref{fig:thermal_ionization}, top row). 

\subsubsection{MRI-active and dead zones} \label{sec:mri_active_dead}
\begin{figure*}
    \centering
	\includegraphics[height=0.26\textwidth,trim={0 0 3.3cm 0},clip]{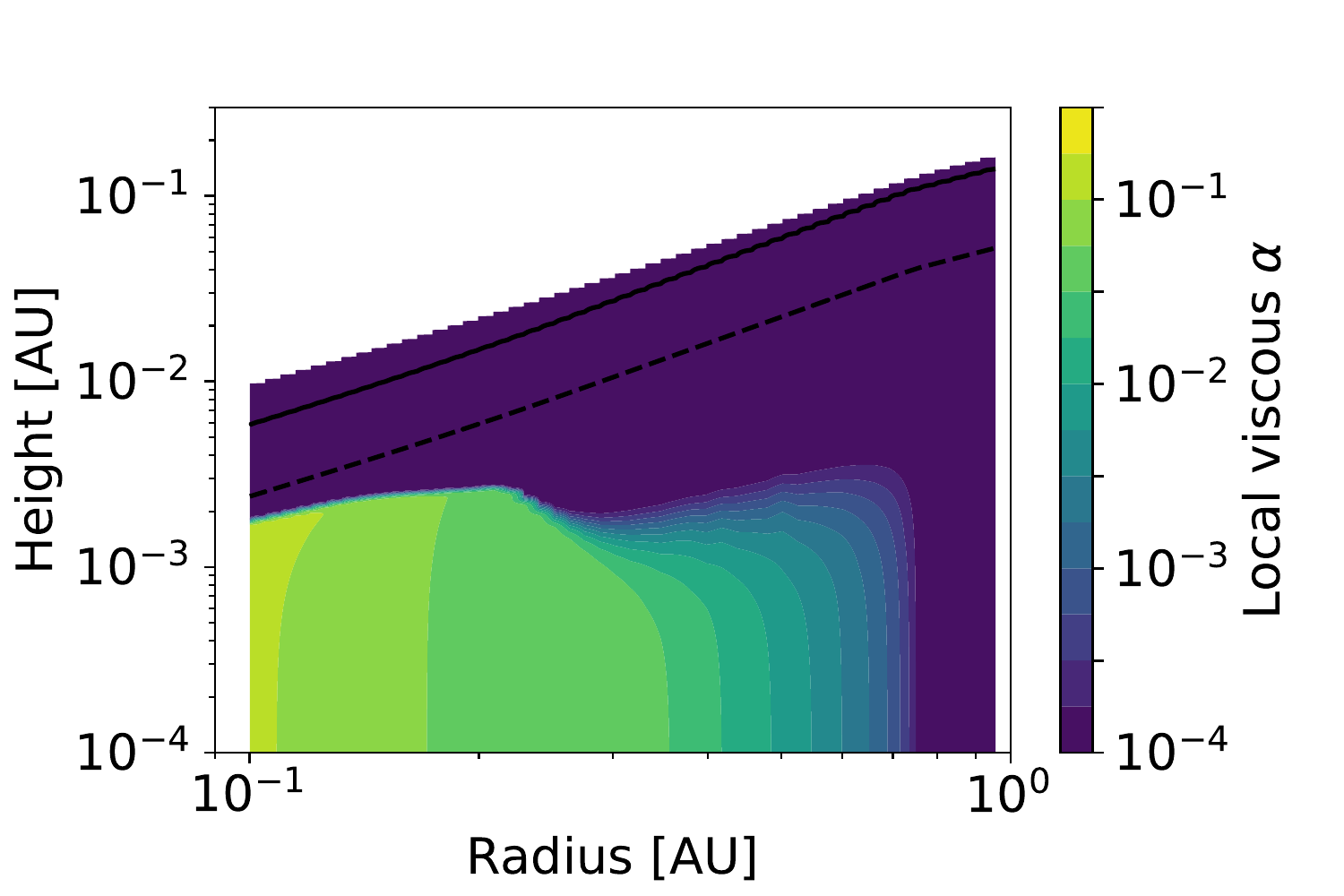}
	\includegraphics[height=0.26\textwidth,trim={2.2cm 0 3.3cm 0},clip]{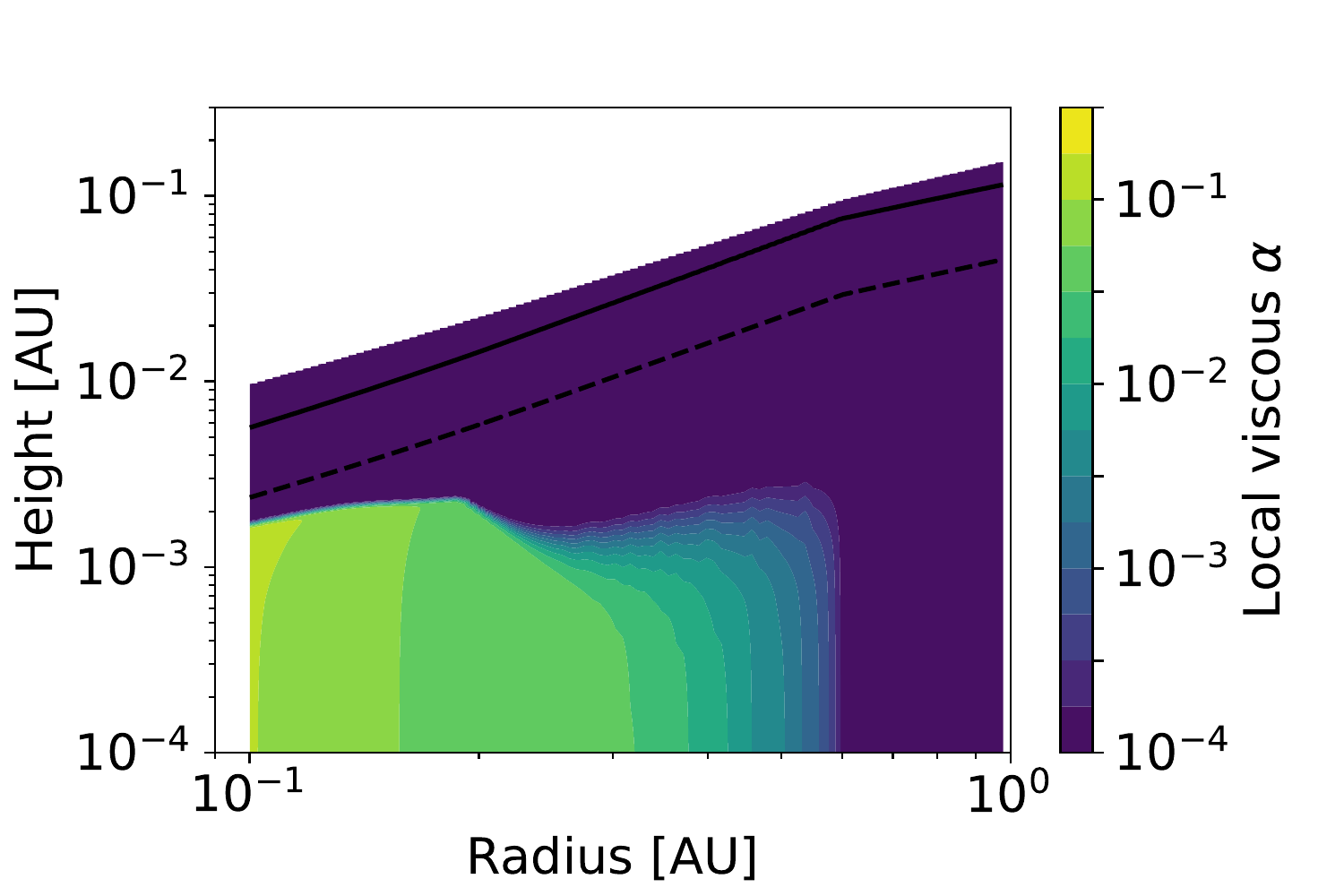}
	\includegraphics[height=0.26\textwidth,trim={2.2cm 0 0 0},clip]{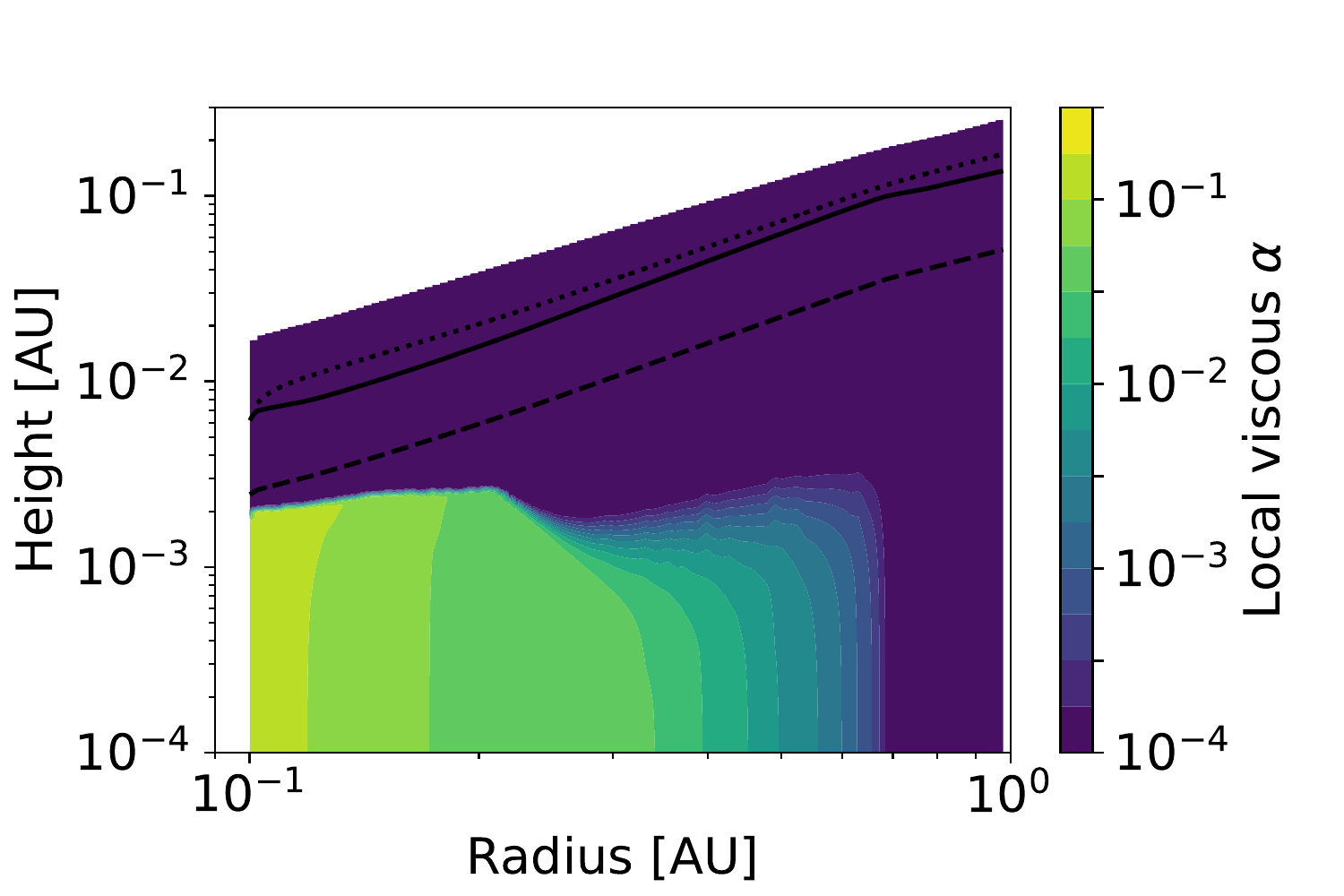}
    \caption{Local viscous parameter $\alpha$ as a function of location in the disc for a viscously-heated constant-opacity model (left), viscously-heated model with realistic opacities (middle), and a viscously and irradiation-heated model with realistic opacities (right). In each panel the solid line shows the disc photosphere ($\tau_{\rm R}=2/3$) and the dashed line shows the pressure scale height ($P=e^{-1/2}P_{\rm mid}$). The dotted line in the right-hand-side panel shows the surface at which $\tau_{\rm irr}=2/3$. Note that the heating by stellar irradiation has a very weak effect on the extent of the MRI-active region. See Section \ref{sec:mri_active_dead}.}
    \label{fig:thermal_alpha}
\end{figure*}

The differences in the ionization structure of our vertically self-consistent models and the vertically isothermal model lead to differences in where the MRI is active in the disc. A plot of the viscosity parameter $\alpha$ as a function of disc radius and height above midplane, Fig. \ref{fig:thermal_alpha}, shows that the MRI-active zone (i.e., where $\alpha>\alpha_{\rm DZ}$) is confined to the vicinity of the midplane in our models. The same was found by \citet{Terquem2008}, who considered a similar vertically self-consistent disc model, for a simple viscosity prescription in which the MRI is activated above a fixed critical temperature. 
This is qualitatively different from the vertically isothermal case, where the MRI-active zone occurs around the midplane in the innermost region, but rises into the upper layers at larger radii \citep[see][]{Mohanty2018}. The latter configuration, where the active zone is sandwiched between a dead zone around the midplane below and an ambipolar-dead zone above, emerges as a consequence of the ionization increasing with height above the disc midplane in the isothermal case, as discussed above. Moreover, comparing our non-irradiated discs (first two panels of Fig. \ref{fig:thermal_alpha}) with the irradiated one (last panel), we see that heating by stellar irradiation makes little difference to the extent of the MRI-active zone, since it only weakly affects the midplane temperature, and the hot uppermost layers are dead due to ambipolar diffusion.

\begin{figure}
    \centering
    \includegraphics[width=0.49\textwidth,trim={0 1.5cm 0 2.5cm},clip]{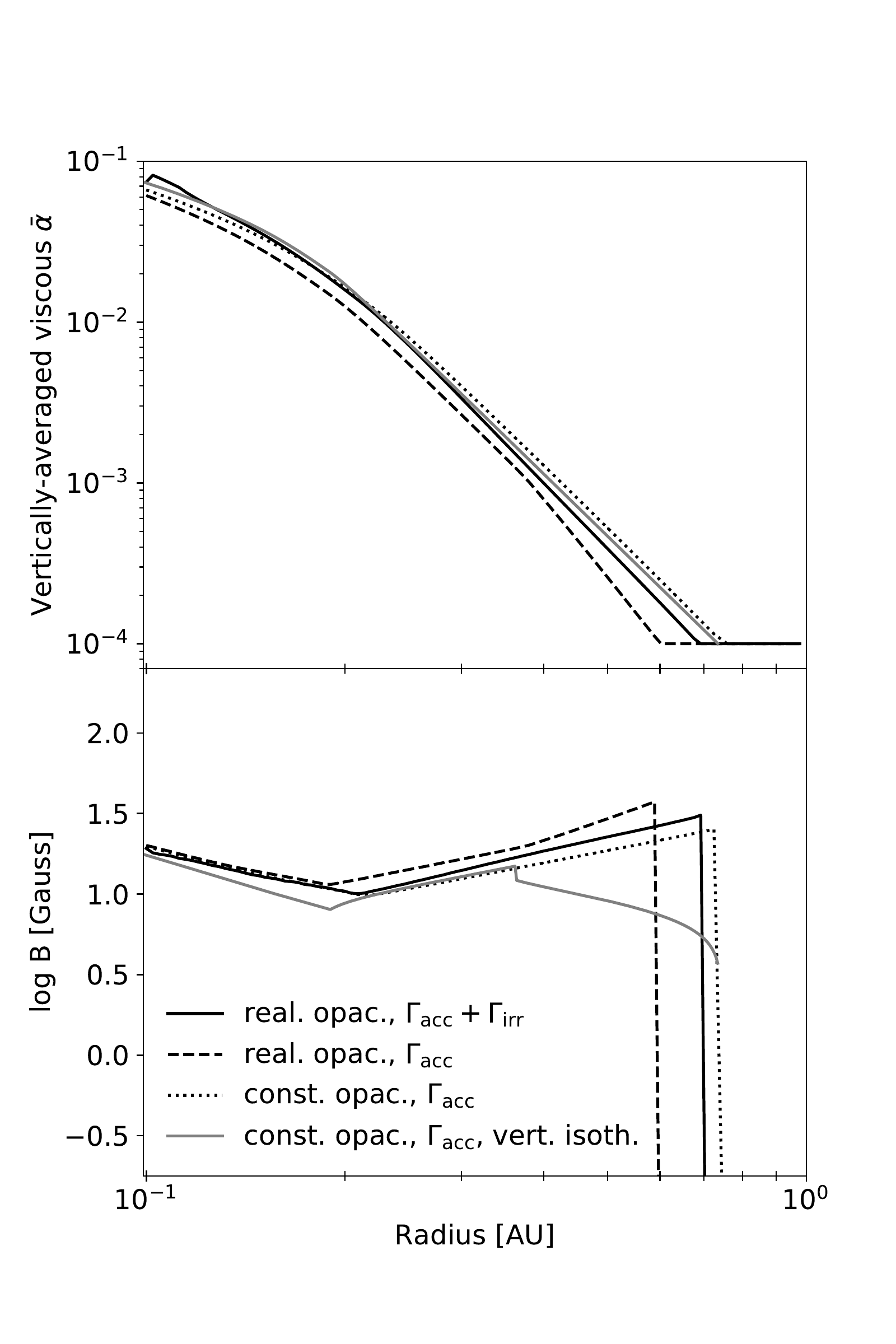}
    \caption{Vertically-averaged viscosity parameter ($\bar{\alpha}$) and the \textcolor{red}{MRI-generated} magnetic field strength ($B$) as functions of radius, in a vertically-isothermal model as well as in self-consistent models of varying complexity (viscously heated constant-opacity model, viscously heated model with realistic opacities, and viscously and irradiation-heated model with realistic opacities), as indicated in plot legend. The vertically-averaged viscosity parameter profile is similar in all four models, but the magnetic field strength profile is qualitatively different in the vertically-isothermal model \textcolor{red}{(in which the field strength decreases as a function of radius outwards from $\sim$ 0.4\,AU)}. See Section \ref{sec:mri_active_dead}.}
    \label{fig:thermal_alphabar}
\end{figure}

Finally, in Fig. \ref{fig:thermal_alphabar} we compare radial profiles of the vertically-averaged viscosity parameter $\bar{\alpha}$ and the \textcolor{red}{MRI-generated} magnetic field strength $B$ from the four models \citep[the vertically isothermal and constant opacity disc of][ plus our three non-isothermal discs]{Mohanty2018}.

In line with theoretical expectations and the results of \citet{Mohanty2018}, the vertically-averaged viscosity parameter $\bar{\alpha}$ decreases as a function of orbital radius. At some distance from the star the MRI is completely quenched and the average viscosity parameter reaches the minimum value, $\bar{\alpha}=\alpha_{\rm DZ}$. Interestingly, the $\bar{\alpha}$ radial profile is both qualitatively and quantitatively similar in all four models. However, the radial profile of the magnetic field strength $B$ reveals qualitative differences. In the vertically-isothermal model, $B(r)$ features a sharp drop at $\sim$\,0.35\,AU, which corresponds to the appearance of a dead zone in the disc midplane, as described above. From that point, until the MRI is completely smothered at $\sim$\,0.7\,AU, the MRI is active in a thin layer high above the midplane, between a dead zone below and an ambipolar-dead zone above \citep[see][]{Mohanty2018}. In the self-consistent models, however, this configuration does not appear: the magnetic field strength remains strong until the MRI is completely quenched, and the MRI-active zone in the inner disc is always restricted to the midplane regions. 


\subsubsection{Surface density and pressure maximum}
\begin{figure}
    \centering
    \includegraphics[width=0.49\textwidth,trim={0 1.5cm 0 2.5cm},clip]{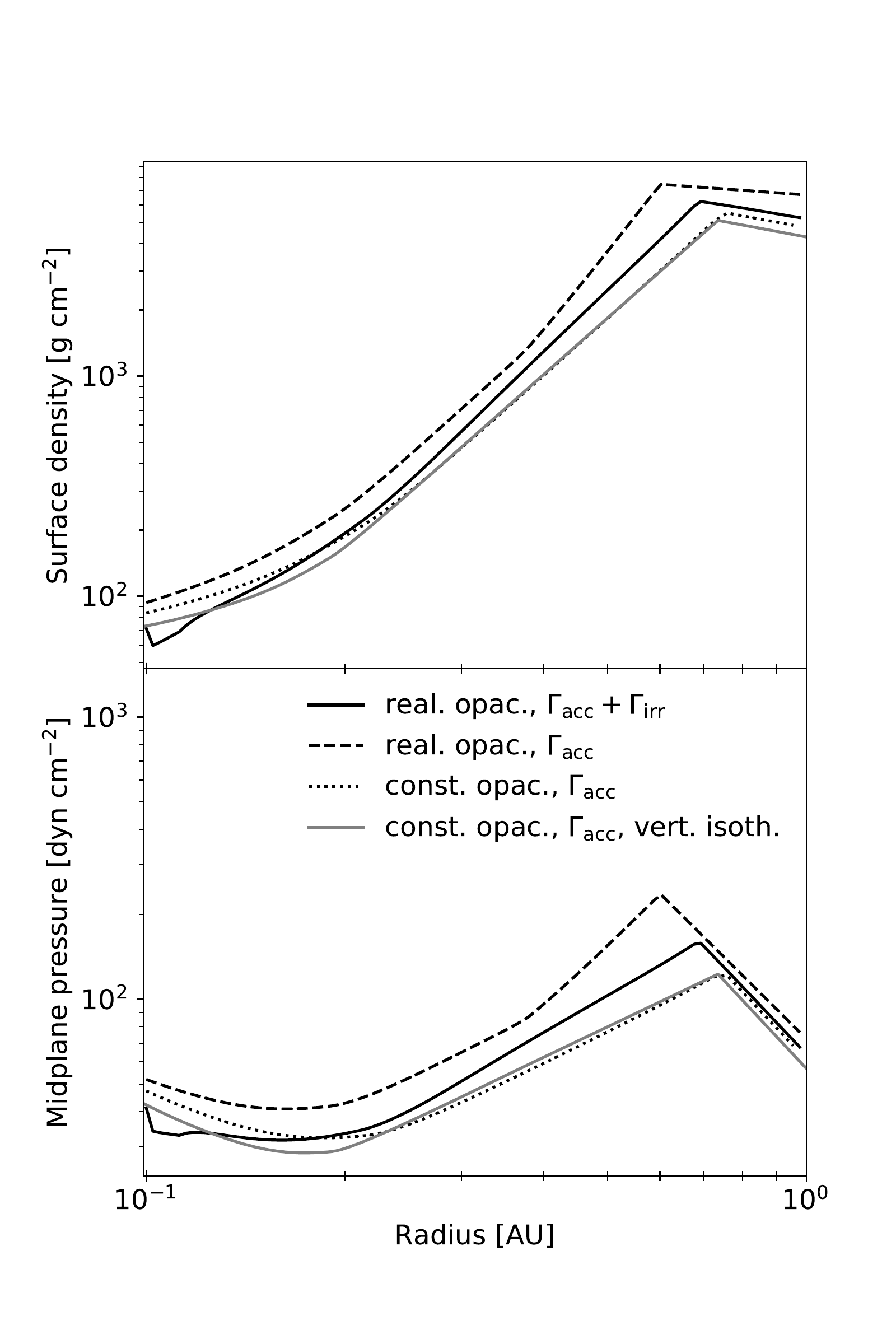}
    \caption{Surface density (top) and midplane pressure (bottom) as functions of radius in a vertically-isothermal and self-consistent models of varying complexity (same as in Fig. \ref{fig:thermal_alphabar}). In each model the density and the midplane pressure maximum correspond to the point at which the viscous $\bar\alpha$ reaches a minimum value. See Section \ref{sec:mri_active_dead}.}
    \label{fig:therm_struct}
\end{figure}

All four steady-state models discussed above feature a maximum in both the local gas pressure and the surface density (Fig. \ref{fig:therm_struct}), at the radial location where the vertically-averaged viscosity parameter $\bar\alpha$ falls to its minimum value (see Fig. \ref{fig:thermal_alphabar}). Inwards of this location, the MRI-driven accretion efficiency (i.e., $\bar\alpha$) increases; for a radially-constant (i.e., steady-state) gas accretion rate, this leads to a decrease in both the surface density and the midplane pressure. Since the radial profile of $\bar\alpha$ is similar in the four models, their surface density and midplane pressure profiles are similar too. Within our three self-consistent models, the addition of heating by stellar irradiation moves the pressure maximum outwards by $\sim$0.1\,AU (compare the dashed and solid black lines in Fig. \ref{fig:therm_struct}, bottom panel). At the same time, relative to the result of \citet{Mohanty2018} for an isothermal and constant opacity disc (grey solid line), the pressure maximum in our most complex model in this section (non-isothermal, with realistic opacities and irradiation heating; black solid line) has moved $\sim$0.1\,AU inwards.

\subsubsection{\textcolor{red}{Summary}}
\textcolor{red}{In this section we have considered how the details of the disc thermal structure model affect the disc's ionization state and where in the disc the MRI is active. We have also examined the importance of the various physical effects on the location of the MRI-induced pressure maximum in the inner disc.}

\textcolor{red}{Our results show that when the disc vertical structure and the MRI-driven viscosity are considered self-consistently, the MRI is active around the disc midplane. In previous \textcolor{red}{work using a simplified vertically-isothermal model, a different} configuration arises in the vicinity of the pressure maximum, where the MRI is active in a layer above (and below) the dead midplane. Though both approaches yield a similar location for the pressure maximum for our fiducial parameters, \textcolor{red}{the difference in the physical behaviour of the disc cautions against the use of the simplified model.}}

\textcolor{red}{Furthermore, for the fiducial parameters considered here, we find that the midplane temperature is primarily determined by viscous dissipation, and not stellar irradiation. As a result, the location of the pressure maximum, where the midplane temperature and the ionization levels fall below a critical value, is negligibly affected by stellar irradiation. Finally, we find that the inner disc is convectively unstable, even if the disc opacity is constant. Hence, even though the location of the pressure maximum \textcolor{red}{is similar to that in previous work}, the disc's structure is qualitatively different. These findings are further discussed in sections \ref{sec:importance_irradiation} and \ref{sec:analytic_convection}.}

\subsection{Disc chemical structure and the MRI} \label{sec:chemical_struct}
In this section we build upon our models from the previous section. We first explore the effects of dust on the disc's ionization state and on the MRI. We then construct our complete model of the inner disc by also considering non-thermal sources of ionization.

\subsubsection{Effects of dust} \label{sec:model_8} 
\begin{figure}
    \centering
    \includegraphics[width=0.49\textwidth,trim={0 4.5cm 0 5.5cm},clip]{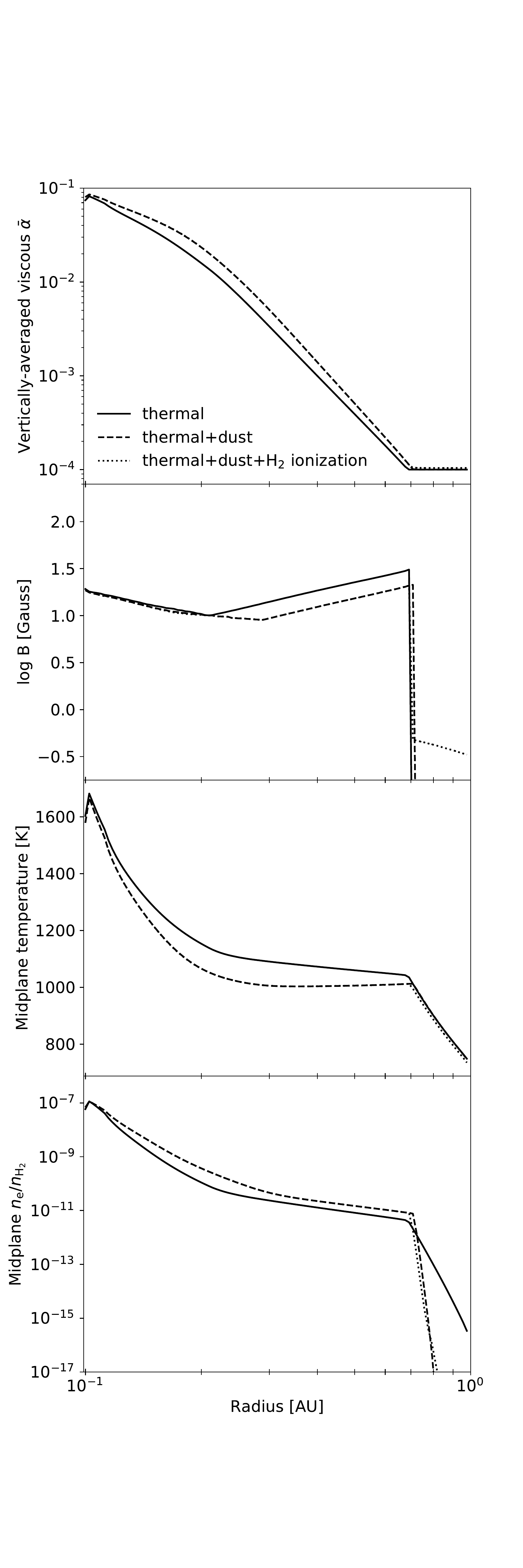}
    \caption{Vertically-averaged viscous parameter ($\bar\alpha$), \textcolor{red}{MRI-generated} magnetic field strength ($B$), midplane temperature and the midplane free electron fraction ($n_{\rm e}/n_{\rm H_2}$) as functions of radius, for a model with thermal ionization only, a model with thermal ionization and dust effects, and a model which also includes realistic non-thermal sources of ionization of H$_2$ ($\zeta=\zeta_{\rm R}+\zeta_{\rm CR}+\zeta_{\rm X}$). The disc structure is quantitatively similar in all three models; however, the main sources of ionization in the models with dust are thermionic and ion emission. See Section \ref{sec:model_8}.}
    \label{fig:chem_comparison}
\end{figure}

\begin{figure}
    \centering
    \includegraphics[width=0.49\textwidth,trim={0 1.5cm 0 2.5cm},clip]{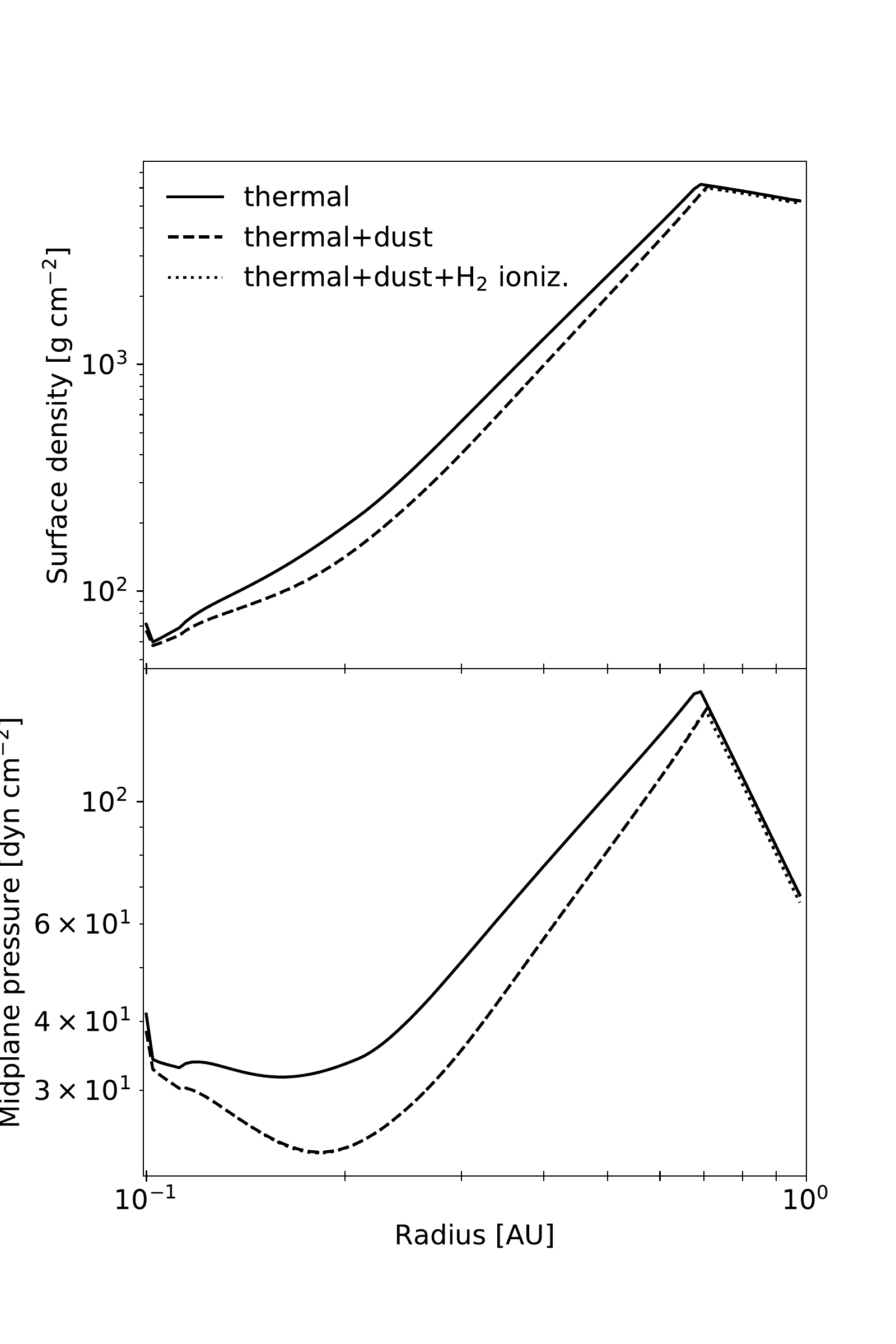}
    \caption{Surface density (top) and midplane pressure (bottom) as functions of radius for a model with thermal ionization only, a model with thermal ionization and dust effects, and a model which also includes realistic non-thermal sources of ionization of H$_2$ ($\zeta=\zeta_{\rm R}+\zeta_{\rm CR}+\zeta_{\rm X}$). The disc structure is  similar in all three. See Section \ref{sec:model_8}.}
    \label{fig:chem_struct}
\end{figure}

Here we consider two models. The first is our most complex model from the previous section, which incorporates a self-consistent vertical structure and realistic opacities, and includes stellar irradiation, but accounts for only gas-phase thermal ionization and recombination of potassium (solid lines in Figs. \ref{fig:chem_comparison} and \ref{fig:chem_struct}). The second is a model including all these processes, as well as dust effects on the ionization (dashed lines in Figs. \ref{fig:chem_comparison} and \ref{fig:chem_struct}); the latter effects include adsorption of neutral atoms and free charges onto dust grains, recombinations on the grain surfaces, and thermionic and ion emission from the grains.

Figure \ref{fig:chem_comparison} depicts the radial profiles of the vertically-averaged viscosity $\bar{\alpha}$, magnetic field strength, midplane temperature and the midplane fractional ionization ($n_{\rm e}/n_{\rm H_2}$) for the two models.  We see that the radial profile of $\bar{\alpha}$, and the radial extent of the MRI-active zone (i.e., where $\bar{\alpha} > \alpha_{\rm DZ}$), are similar in the two models. Consequently, their surface density and midplane pressure profiles, plotted in Fig. \ref{fig:chem_struct}, are also similar. Interestingly, Fig. \ref{fig:chem_comparison} also shows that when the influence of dust on the ionization is included, both $\bar{\alpha}$ and the ionization fraction are somewhat higher at a given radius within the active region, while the midplane temperature is lower. We discuss these effects further in section \ref{sec:effects_dust}. 

\begin{figure*}
    \centering
    \includegraphics[width=\textwidth,trim={0 0 0 0}, clip]{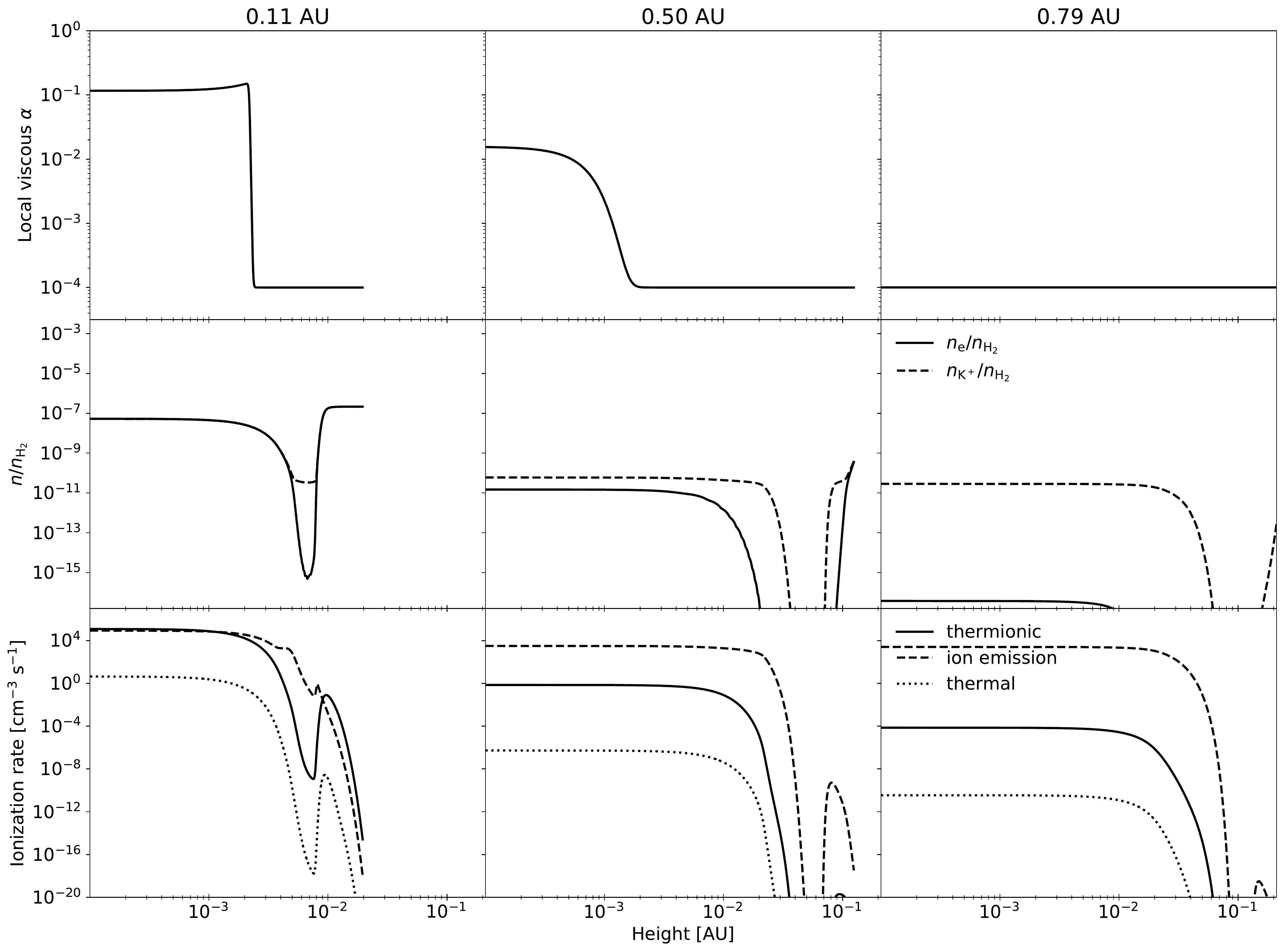}
    \caption{Local viscous parameter $\alpha$ (top), fractional abundance of charged species $n/n_{\rm H_2}$ (middle) and ionization rates (bottom; thermionic $\mathcal{R}_{\rm therm}$, \textcolor{red}{ion emission $\mathcal{R}_{\rm K, evap} f_+$,} thermal $k_2 n_{\rm H_2} n_{\rm K^0}$) at three different radii (as indicated in panel titles) for the model with thermal ionization and dust effects. Thermionic emission is the primary source of free electrons in the MRI-active regions. See Section \ref{sec:model_8}.}
    \label{fig:chem_8}
\end{figure*}

Figure \ref{fig:chem_8} illustrates the vertical structure of the model with dust at three different disc radii. The top row shows the local viscosity parameter $\alpha$, the middle row shows the number densities of free electrons and ions, and the bottom row shows the various contributions to the free electron production rate per unit volume. As in the thermally ionized disc (Fig. \ref{fig:thermal_ionization}), the MRI is active in the hot disc midplane. The ionization levels decrease with height above the midplane as the temperature decreases, and increase again in the disc atmosphere heated by stellar irradiation. In the hot upper layers ambipolar diffusion quenches the MRI.

The plots of free electron and ion production rates show that thermionic and ion emission are the dominant ionization sources. Thus, Fig. \ref{fig:chem_comparison} is misleading in the sense that, while the differences in the global structure are very small when dust effects are added, {\it it is not because dust effects are minor}. Clearly, dust dominates the chemistry in the inner disc. Rather, for the parameters assumed here, the ionization levels as a function of temperature and density, when dust is included, are similar to the levels obtained from gas-phase thermal ionization only, {\it due to the similarity between the ionization potential of potassium and the grain work function}.

\subsubsection{Non-thermal sources of ionization} \label{sec:model_9}
In this section we present our full model of the inner disc which, in addition to the above dust effects, also includes non-thermal sources of ionization: stellar X-rays, cosmic rays and radionuclides all ionize H$_2$, ultimately producing metallic ions and free electrons. The resulting radial profiles of $\bar{\alpha}$, magnetic field strength, midplane temperature, density and ionization levels are shown in Fig. \ref{fig:chem_comparison} as dotted lines. The radial profile of $\bar{\alpha}$ seems to overlap with the model with only dust effects (dashed line). However, in the region where the MRI is dead in the dust-only model, $\bar{\alpha}$ is slightly higher than $\alpha_{\rm DZ}$ in the new model including H$_2$ ionization. The magnetic field strength also does not drop to zero in this model, further revealing that the MRI remains active here.

\begin{figure*}
    \centering
    \includegraphics[width=\textwidth,trim={0 0 0 0}, clip]{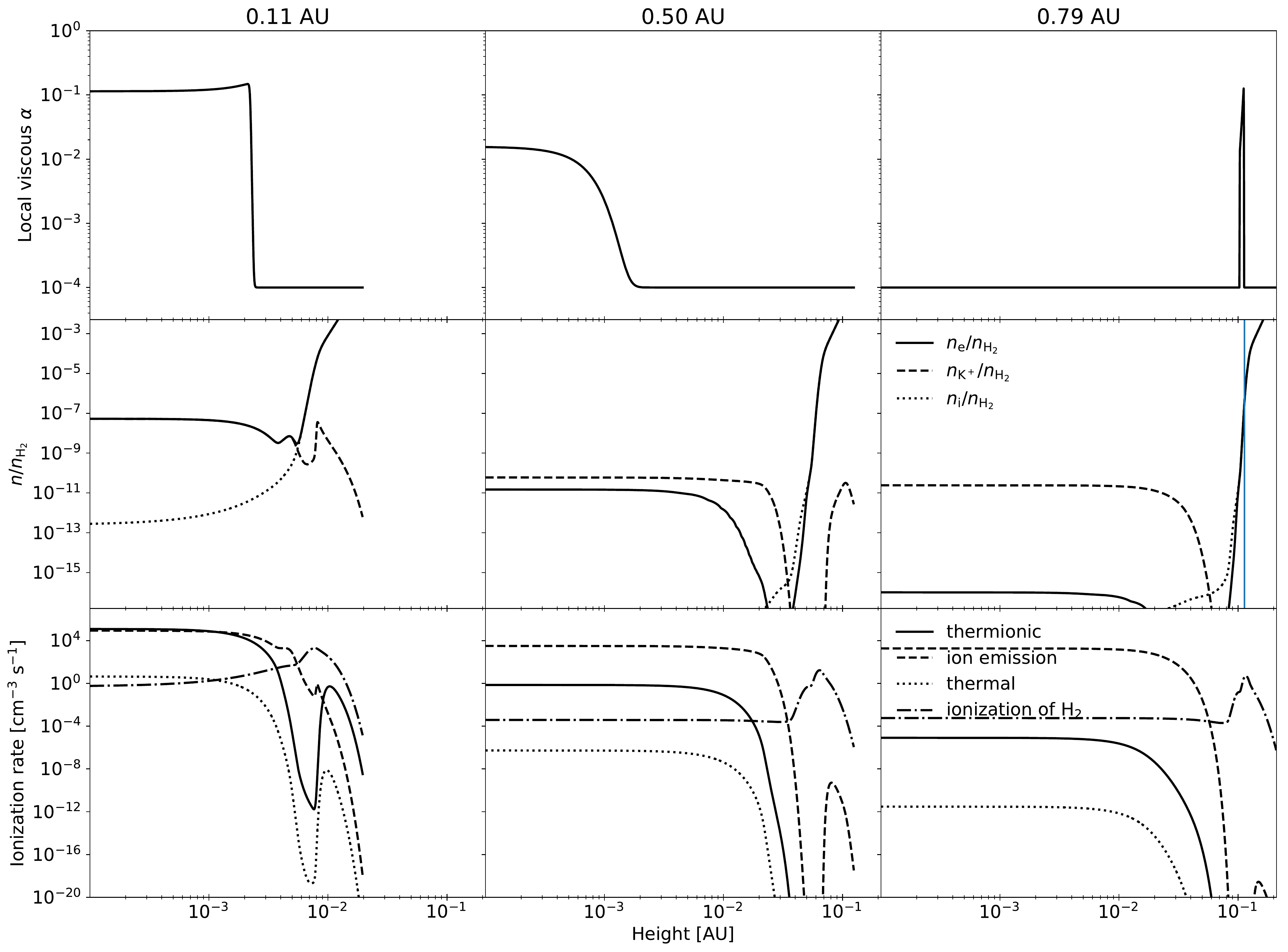}
    \caption{Local viscous parameter $\alpha$ (top), ionization fraction \textcolor{red}{(middle; electrons: $n_{\rm e}/n_{\rm H_2}$, potassium ions: $n_{\rm K^+}/n_{\rm H_2}$, metal ions: $n_{\rm i}/n_{\rm H_2}$)} and ionization rates (bottom; thermionic $\mathcal{R}_{\rm therm}$, \textcolor{red}{ion emission $\mathcal{R}_{\rm K, evap} f_+$,} thermal $k_2 n_{\rm H_2} n_{\rm K^0}$ and non-thermal $\zeta n_{\rm H_2}$) at three different radii (as indicated in panel titles) for the model with all sources of thermal and non-thermal ionization. Non-thermal ionization produces an MRI-active region high above disc midplane at larger radii (see the top right panel). The blue line in the right panel on the second row indicates the upper boundary of this MRI-active region at the given radius, and the ionization fraction at which it occurs. \textcolor{red}{Note that the metal ion fraction becomes unrealistically large only to the right of the blue line, i.e., only above the high-altitude MRI-active layer}. See Section \ref{sec:model_9}.}
    \label{fig:chem_9}
\end{figure*}

\begin{figure}
    \centering
	\includegraphics[width=0.49\textwidth,trim={0 0 0 0},clip]{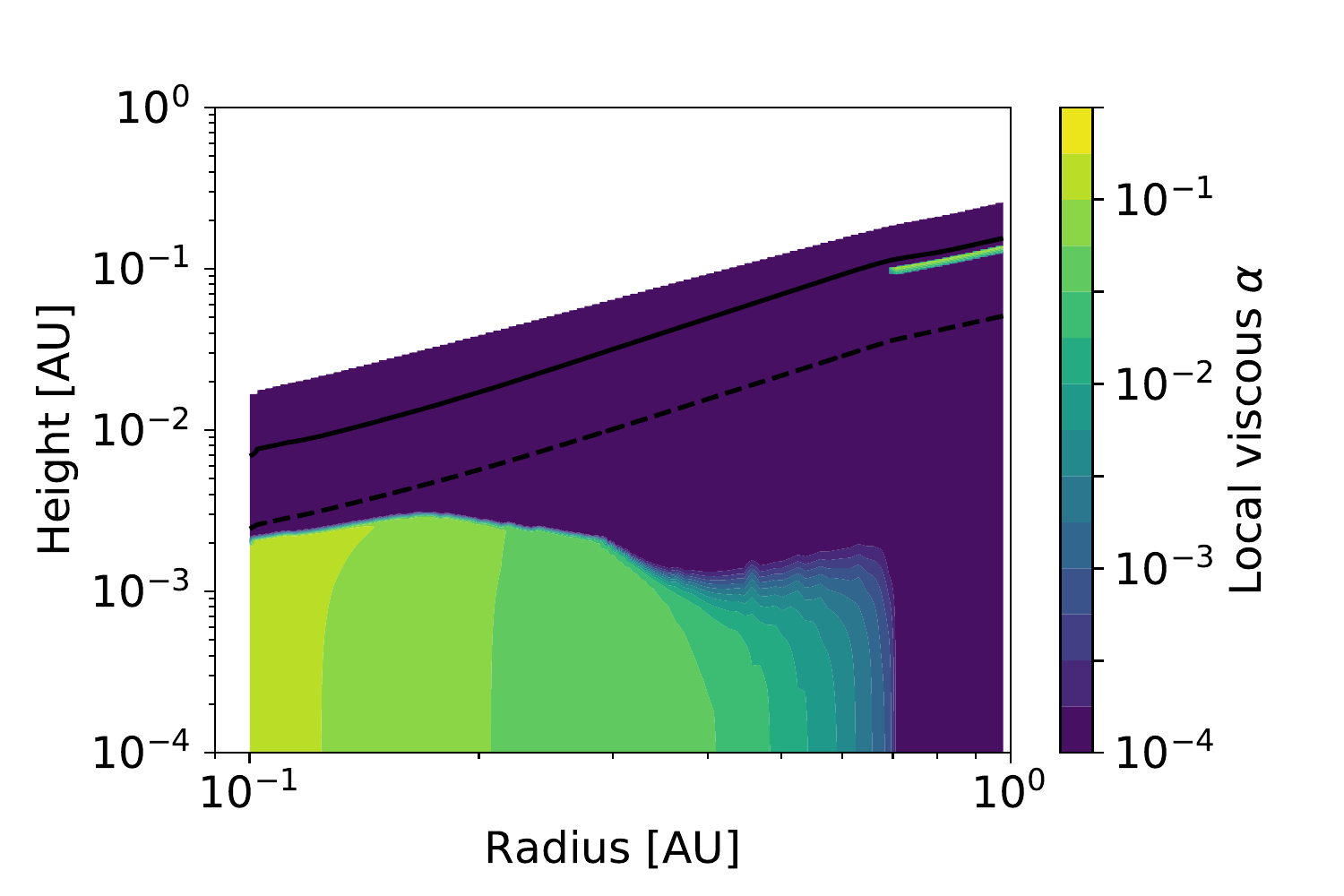}
    \caption{Local viscous parameter $\alpha$ as a function of location in the disc for the model with all sources of thermal and non-thermal ionization. In the innermost disc, thermionic and ion emission ionize the dense regions around the disc midplane, producing the high MRI-driven $\alpha$ there. At larger radii, the MRI is active in a \textcolor{red}{low-density} layer high above the disc midplane, dominated by non-thermal sources of ionization. The solid black line indicates the surface in the disc above which the ionization fraction $n_{\rm e}/n_{\rm H_2}>10^{-4}$, i.e., above which the assumptions of our simple chemical network break; this surface is above the MRI-active region at all radii. See Section \ref{sec:model_9}.}
    \label{fig:chem_9_alpha}
\end{figure}

Figure \ref{fig:chem_9} shows the vertical structure of this disc at three different radii. It reveals that, at large radii where the MRI was quenched in previous models, the MRI now remains active at large heights (see also Fig. \ref{fig:chem_9_alpha}). 
The situation is reminiscent of the appearance of an MRI-active layer near the disc surface in the thermally ionized, vertically isothermal model of \citet{Mohanty2018}, but the physical reason is very different: in the latter model, it is due to the isothermal assumption, which leads to increasing fractional ionization with height; in our present non-isothermal model, it is due to additional, non-thermal sources of ionization (mainly X-rays; see below) which elevate the fractional ionization near the disc surface.  

The plot of ionization levels in Fig. \ref{fig:chem_9} shows that the fractional abundances of free electrons and metal ions increase strongly towards the disc surface, where non-thermal ionization dominates. As a result, potassium ions are depleted in the upper layers via recombination with the abundant electrons. Furthermore, the values of the metal ion abundance near the surface are themselves noteworthy: they greatly exceed the solar abundance of Mg, and even of C and O. The reason is as follows. In our chemical network, the ionization of an H$_2$ molecule by a non-thermal source effectively produces a free electron and a metal ion, the latter through (implicitly included) rapid charge exchange between a metal atom and the H$_2$ ion. This is valid as long as the number of H$_2$ ions remains lower than the total number of metal atoms; in this case, the precise total abundance of metals is unimportant, and is not accounted for in our calculations. When the H$_2$ ion abundance exceeds that of metal atoms, however, our network fails: it yields a spuriously high metal ion abundance, when in reality H$_2$ ions dominate (since there are no remaining metal atoms to transfer their charge to). Additionally, when the H$_2$ ion abundance becomes very high (e.g., when $n_{\rm H_2^+}/n_{\rm H_2}$ exceeds $\sim$10$^{-3}$), our assumption that H$_2$ is effectively a neutral species due to charge transfer also breaks down.    


Clearly, above this level our chemical model is not applicable, as the ionized hydrogen would become an important species and our assumption of a constant hydrogen number density would be invalid. Nevertheless, this is above the MRI active zone at all orbital radii (see the black solid line in Fig.\,\ref{fig:chem_9_alpha}, \textcolor{red}{and also the blue solid line in Fig.\,\ref{fig:chem_9}}), and is thus not germane to our conclusions.

\begin{figure*}
    \centering
    \includegraphics[width=\textwidth,trim={0 0 0 0}, clip]{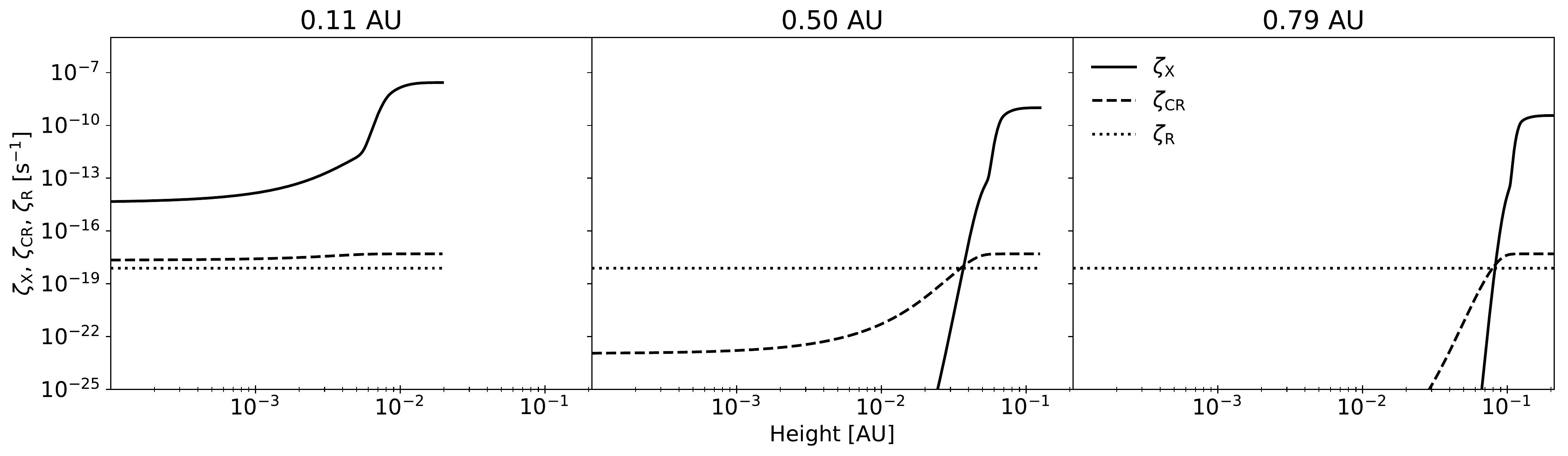}
    \caption{Ionization rates of molecular hydrogen due to stellar X-rays ($\zeta_{\rm X}$), cosmic rays ($\zeta_{\rm CR}$), and radionuclides ($\zeta_{\rm R}$) for the model with all sources of thermal and non-thermal ionization. Stellar X-rays are the dominant source of ionization in the disc upper layers. See Section \ref{sec:model_9}.}
    \label{fig:chem_9_zeta}
\end{figure*}

Figure \ref{fig:chem_9_zeta} compares the contributions from stellar X-rays and cosmic rays to the hydrogen ionization rate. As expected, the un-attenuated ionization rate due to X-rays is higher at the disc surface, but cosmic rays can penetrate deeper in the disc. Nevertheless, the MRI-active region in the upper disc layers is mostly ionized by X-rays, in agreement with previous work \citep[e.g.][]{Glassgold1997,Ercolano2013}.

Finally, note that in this outer region, \textcolor{red}{where the MRI is active in the upper disc}, the gas accretes primarily through the dead zone \textcolor{red}{(driven by the dead-zone viscosity $\alpha_{\rm DZ}$)}, since the density at the dead disc midplane is much higher than in the X-ray-ionized MRI-active layer. This is why the vertically-averaged viscosity parameter is close to the dead-zone value, $\bar\alpha \sim \alpha_{\rm DZ}$, outwards from the pressure maximum.

\subsubsection{Multiple solutions for the vertical disc structure} \label{sec:multiple_solutions}
In the results presented so far, for our fiducial choice of disc and dust parameters, solutions for the vertical disc structure (i.e., solutions in disc height $z_{\rm surf}$) appear to be unique. In general, there is also a single peak in the vertically-averaged viscosity $\bar\alpha$ as a function of magnetic field strength $B$ (which determines our choice for $B$; see section \ref{sec:magnetic_field}). The exception is the vicinity of the orbital radius at which the MRI is quenched at the disc midplane (see Fig. \ref{fig:chem_9_alpha}). There, $\bar\alpha(B)$ has two peaks, one corresponding to the solution where the MRI is active at disc midplane, and the other to the solution where the MRI is active in the upper disc layers, mostly ionized by stellar X-rays. As discussed in section \ref{sec:numerical_methods}, we choose $B$ such that $\bar\alpha$ is maximized in this case as well.

Note that there could be, in principle, an MRI-active layer high up in the disc at shorter orbital radii as well, in addition to the active layer at midplane. Here, this does not appear due to our assumption that the magnetic field strength $B$ is vertically constant. At the high $B$ necessary to drive efficient accretion at midplane, the MRI is quenched in the low-density disc atmosphere due to ambipolar diffusion (since the ambipolar criterion for active MRI also encapsulates the requirement that the magnetic pressure should be less than the thermal pressure). Therefore, it is only when the temperature drops and high-temperature ionization \textcolor{red}{effects (thermal ionisation of potassium, thermionic and ion emission from grains) can no longer} drive the accretion efficiently \textcolor{red}{in the midplane} that our model features the active layer \textcolor{red}{(generated by X-ray ionisation)} high above the disc midplane.

For a different choice of parameters, e.g. if the maximum dust grain size is $a_{\rm max}=100$\,$\mu$m, there may exist an additional range of orbital radii where there are multiple peaks in $\bar\alpha(B)$ and also multiple solutions for the disc vertical structure (for $z_{\rm surf}$) at a fixed value of the magnetic field strength $B$. Similarly to the case above, this issue arises due to the competing effects of high-temperature sources of ionization and X-rays. As these sources of free electrons \textcolor{red}{have different dependencies} on the disc structure (density, temperature, column density), their combination leads to non-monotonous variations in the electron number density as a function of height above the disc's midplane. Since the viscous dissipation due to the MRI is sensitive to the ionization fraction, the total dissipation can be a non-monotonous function of $z_{\rm surf}$. Since the solution for the vertical disc structure is determined by an equilibrium between an input and an output total heat, this can lead to multiple solutions in $z_{\rm surf}$.

\begin{figure}
    \centering
    \includegraphics[width=0.49\textwidth,trim={0 2.5cm 0 2cm}, clip]{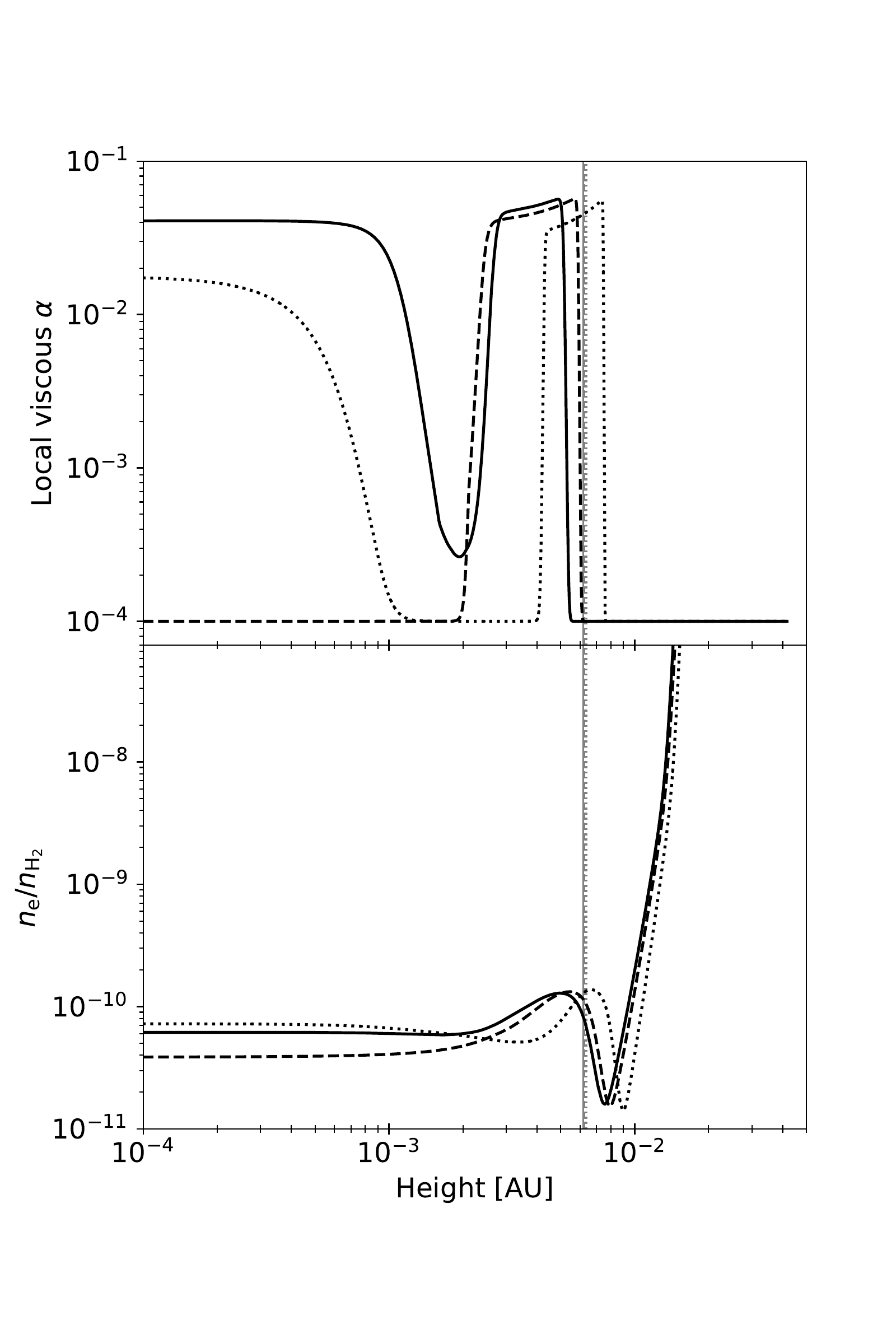}
    \caption{Degeneracy in the vertical disc structure at a fixed magnetic field strength in a model where the maximum dust grain size is $a_{\rm max}=100$\,$\mu$m: 
    local viscous parameter $\alpha$ as a function of height (top), and the fractional electron number density ($n_{\rm e}/n_{\rm H_2}$; bottom) as functions of height in three different equilibrium solutions. The different solutions arise from vertical variations in the viscous $\alpha$; the lengthscales of these variations are much smaller than the disc pressure scale height (shown for each solution by the vertical gray lines). The solutions shown here are at a radius of 0.22\,AU, \textcolor{red}{for field strength} $\textrm{log\,}B_z = 0.226$ and \textcolor{red}{grazing angle} $\phi=0.06$. \textcolor{red}{The vertically-averaged viscous $\bar\alpha$ for the solid, dashed and dotted \textcolor{red}{lines are}, respectively, $2 \times 10^{-2}$, $1.78 \times 10^{-2}$, and $1.28 \times 10^{-2}$.} See Section \ref{sec:multiple_solutions}.}
    \label{fig:multiple_solutions}
\end{figure}

To illustrate this issue, we show in Fig. \ref{fig:multiple_solutions} an example of three thermally-stable solutions for the vertical disc structure at a fixed value of magnetic field strength $B$ that appears in our model for a maximum grain size $a_{\rm max}=100$\,$\mu$m (we ignore thermally unstable solutions). Note that the dependence of the overall disc structure and the location of the pressure maximum on the dust grain size is presented and thoroughly discussed in our companion paper (Jankovic et al. {\it in prep.}). Here we only discuss how we deal with the multiple solutions. Fig. \ref{fig:multiple_solutions} shows the viscosity $\alpha$ as a function of height in the top panel and the ratio $n_{\rm e}/n_{\rm H_2}$ in the bottom.

Evidently, small variations in the free electron number density correspond to large variations in the viscosity $\alpha$, all at heights below one disc pressure scale height (indicated by gray lines). This implies that the difference between these solutions is likely unphysical for two reasons. First, the viscosity $\alpha$ is in reality driven by turbulence, and turbulent motions should not abruptly change over length-scales much smaller than a single pressure scale height. Second, chemical species can also be expected to be spatially mixed by turbulence, and so such vertical variations in the ionization fraction as we obtain here might be smoothed over in reality. Since resolving these issues is beyond the scope of our models, we simply always choose a solution with minimum $z_{\rm surf}$, which also appears to always correspond to a maximum $\bar\alpha$ at the given magnetic field strength. 

\subsubsection{\textcolor{red}{Summary}}
\textcolor{red}{As previously suggested by \citet{Desch2015}, the inner disc is primarily ionized through thermionic and potassium ion emission from dust grains. These processes counteract adsorption of free charges onto dust grains at the high temperatures present in the inner disc. We show that, for our fiducial parameters, the introduction of dust effects on disc ionization has very little effect on the location of the pressure maximum. This is because thermionic and ion emission become efficient above roughly the same threshold temperature as thermal ionization of potassium. Additionally, we show that non-thermal sources of ionization are unimportant for the fiducial parameters considered here. Thus, similar to our results in Section 3.1, while the inclusion of new physics \textcolor{red}{in} our model has not fundamentally changed the position of the pressure maximum, the physics setting its position is \textcolor{red}{very different from} that described in \citet{Mohanty2018}.  }

\section{Discussion} \label{sec:discussion}

\subsection{Effects of dust}
\label{sec:effects_dust}
A key feature of the inner disc is that the MRI drives high viscosity in the innermost regions, close to the star, but becomes largely suppressed at larger orbital distances (in the so-called dead zone). This leads to the formation of a local gas pressure maximum that may play a key role in planet formation at short orbital distances \citep{Chatterjee2014}. This decrease in viscosity is expected to arise because the innermost regions are hot enough (>1000\,K) to thermally ionize potassium (coupling the gas to the magnetic field), but further out temperature and ionization levels decrease substantially \citep{Gammie1996}. In a previous study, we showed that in a thermally-ionized disc coupled self-consistently to an MRI viscosity, the inner edge of the dead zone lies at a few tenths of an AU \citep{Mohanty2018}.

One of the key differences \textcolor{red}{between this work and that of} \citet{Mohanty2018} is that here we also take into account the effects of dust on the disc's ionization state. Small dust grains present in the disc sweep up free electrons and ions from the gas, and these recombine quickly on the grain surfaces. In the bulk of the protoplanetary disc dust grains therefore efficiently lower the ionization fraction, decoupling the magnetic field from the gas and suppressing the MRI \citep{Sano2000, Ilgner2006, Wardle2007, Salmeron2008, Bai2009, Mohanty2013}. However, in the inner regions of protoplanetary discs dust grains also act to increase the ionization levels, as at high temperatures they can also emit electrons and ions into the gas \citep{Desch2015}. The balance between thermal ionization and these processes then determines how well ionized the inner disc is, and thus the extent of the high-viscosity region and the location of the dead zone inner edge.

The top panel of Fig. \ref{fig:chem_comparison} shows that addition of dust only weakly affects the vertically-averaged viscosity $\bar\alpha$ in the inner disc. At a given orbital radius, $\bar\alpha$ is even slightly higher than in the model with no dust, implying that thermionic and ion emission are important sources of ionization. In fact, as \citet{Desch2015} showed, thermionic and ion emission can become the main source of free electrons at high temperatures. For our disc model this can be seen in the bottom panel of Fig. \ref{fig:chem_9}, which shows that at the hot disc midplane thermionic and ion emission dominate over other sources of ionization. Clearly then, for the chosen parameters, the adsorption of charges onto grains is more than offset by the expulsion of charges from hot grain surfaces.

Similarity in the resulting disc structure in the models with and without dust grains can be explained by the similar dependency on temperature that thermal and thermionic/ion emission have (and which can also be deduced from the bottom panel of Fig. \ref{fig:chem_9}). As discussed by \citet{Desch2015}, thermal ionization of potassium becomes efficient at temperatures above $\sim 1000$\,K in accordance with its ionization potential $\textrm{IP}=4.34$\,eV. The temperature at which thermionic and ion emission become important is determined by the work function $W$ of the material out which dust grains are made, and for silicates $W \sim 5$\,eV. This alone would imply that thermionic emission becomes efficient at temperatures closer to $2000$\,K. However, above $\sim 1000$\,K potassium-bearing minerals start evaporating from grain surfaces \citep{Lodders2003}, and a fraction of potassium atoms leaves the grain surface as ions (since $W \sim \textrm{IP}$; \textcolor{red}{see eq.} (\ref{eq:pot_ion_atom_ratio})). Dust grains then become negatively charged, which reduces the effective potential that electrons need to overcome for thermionic emission (the effective work function $W_{\rm eff}$).

The above results could change significantly as dust grains grow or as they accumulate in the inner disc (as needed for the formation of solid planet cores). Dust adsorption of free charges, for example, becomes much less efficient for larger grains, since the total grain surface area decreases \citep{Sano2000, Ilgner2006}. We investigate how dust growth and varying dust-to-gas ratio affect the inner disc structure in our companion paper (Jankovic et al. {\it in prep.}).

\subsection{\textcolor{red}{Effect} of stellar irradiation} \label{sec:importance_irradiation}
We find that the absorbed flux of stellar irradiation is many times higher than the heat flux generated by accretion at any given radius, yet irradiation has a very small effect on the disc midplane temperature. Consequently, the ionization levels and the MRI-driven viscosity are similar in the models with and without stellar irradiation. \textcolor{red}{Why does the irradiation have such a marginal effect?}

Essentially, it is because the stellar irradiation heats the disc's optically-thin regions, from which heat escapes easily. Accretion heat is generated deep in the disc, where the optical depth is much higher. In the absence of stellar irradiation, the midplane temperature in the optically thick disc is $\sigma_{\rm SB} T_{\rm mid}^4 \sim F_{\rm acc} \tau_{\rm mid}$ \citep[][though in our work the midplane temperature is somewhat lower due to convection]{Hubeny1990}. If on top of a viscously-heated layer of optical thickness $\tau_{\rm mid} \gg 1$ there is an irradiation-heated layer of optical thickness $\tau_{\rm upper}$, \textcolor{red}{it follows from eq. (\ref{eq:rt_2}) and (\ref{eq:rt_3})} that $\sigma_{\rm SB} T_{\rm mid}^4 \sim F_{\rm acc} \tau_{\rm mid} + F_{\rm irr} \tau_{\rm upper}$ (again, neglecting convection). Here $\tau_{\rm upper}$ is the optical depth of the disc to its own radiation down to a height at which the disc becomes optically thick to stellar irradiation. Then, if $\tau_{\rm mid}$ is sufficiently larger than $\tau_{\rm upper}$, the midplane temperature is determined by viscous dissipation.

Our results are consistent with those of \citet{DAlessio1998}, who also found that models with and without stellar irradiation yield roughly the same midplane temperatures in the optically-thick inner disc. Similarly, \citet{Flock2019} considered 2D static radiation-hydrodynamics models of the inner disc heated by stellar irradiation only, but found that the midplane temperature (and the orbital radius of the dead zone inner edge) would increase appreciably if accretion heat were included, on the condition that the accretion heat is released near the optically-thick midplane.

Note, additionally, that we have not considered the details of the inner disc edge or the dust sublimation line. If the inner rim of the disc is puffed-up, it would throw a shadow over a portion of the inner disc \citep{Natta2001, Dullemond2001}, further reducing the importance of stellar irradiation.

\subsection{Convective instability in the inner disc} \label{sec:analytic_convection}
In section \ref{sec:thermal_struct} we showed that a large region of the inner disc is convectively unstable. We find this to be the case even when the opacities are constant, i.e., a super-linear growth of the opacity with temperature \citep{Lin1980} is not needed. Here, the high temperature gradient is established because the heat is deposited deep within the optically-thick disc. In the presented models most of the viscous dissipation happens near the midplane, where the MRI is active, but the same is also true for a vertically-constant viscosity $\alpha$. This result can also be confirmed analytically. We consider a simplified problem of radiative transfer in the optically thick limit, where the temperature is given by $\sigma T^4 = \frac{3}{4} \tau F(z_{\rm surf})$ \citep{Hubeny1990}. Assuming a constant disc opacity, the equation of hydrostatic equilibrium can be re-written as $\frac{dP}{d\tau} = \frac{\Omega^2 z}{\kappa_{\rm R}}$. The temperature gradient is then given by
$$ \nabla = \frac{d \textrm{ln} T}{d \textrm{ln} P} = \frac{\kappa_{\rm R} P}{4 \Omega^2 z \tau} . $$
The appropriate upper boundary condition for this problem is the disc photosphere ($\tau=2/3$), where the gas pressure is given by $P_{\rm surf}=\Omega^2 z_{\rm surf} \tau_{\rm surf} / \kappa_{\rm R}$ \citep[assuming that the disc is vertically isothermal above the photosphere,][]{Papaloizou1999}. At the photosphere, given the chosen boundary condition, we have $\nabla = 1/4$. Near the midplane the optical depth is $\tau_{\rm MID}=\frac{1}{2}\kappa_{\rm R} \Sigma$, and we may estimate the midplane pressure as
$$ P_{\rm MID} = \rho_{\rm MID} c_{\rm s, MID}\textcolor{red}{^2} = \frac{\Sigma}{2 H} c_{\rm s, MID}\textcolor{red}{^2} = \frac{1}{2} \Omega^2 \Sigma H , $$
where the disc's scale height $H$ is related to the midplane temperature through hydrostatic equilibrium. Substituting $\tau_{\rm MID}$ and $P_{\rm MID}$ into the expression for the temperature gradient, we have
$$ \nabla_{\rm MID} = \frac{1}{4} \frac{H}{z} , $$
implying that such a disc should become convectively unstable a bit below one scale height. However, we can further estimate the gradient $\nabla$ near one scale height, by assuming that there $P_{H} \sim P_{\rm MID} e^{-1/2} \sim 0.6 P_{\rm MID}$, and $\tau_{H} \sim 0.3 \tau_{\rm MID}$, as follows from vertically isothermal, Gaussian profiles of pressure and density. Thus, near $z=H$, we have
$$ \nabla_{H} = \frac{1}{2} \frac{H}{z} , $$
showing that the disc should become convectively unstable above one scale height. 

Furthermore, in the convectively unstable regions we use a simple approximation that convection is efficient and the temperature gradient is isentropic. More detailed calculations would yield an answer in which the temperature gradient lies between the isentropic one and the gradient given by the radiative transport. As it turns out, the result would not differ much from what is obtained here. In the optically-thick limit considered above, the difference in the temperature profile when the entire flux is transported by radiation and when the entire flux is transported by convection is a very weak function of optical depth, and remains small for rather large optical depths \citep{Cassen1993}.

Limiting the temperature gradient is the only role of convection in our simple viscous model. In real discs, convection might interact with the MHD turbulence induced by the MRI. For example, in simulations of ideal MHD, \citet{Bodo2013} and \citet{Hirose2014} find that convection can increase the angular momentum transport driven by the MRI by increasing the magnetic field strength, although it appears that this is only the case when convection is particularly strong \citep{Hirose2015}. Concurrently, it is found that the relationship between the induced stress and the magnetic field strength are not modified. The consequences for the non-ideal MHD regime, relevant in protoplanetary discs, are not clear. In our solutions the value of the vertically-averaged MRI-driven viscosity would decrease with both a decrease and an increase in the magnetic field strength due to non-ideal effects, as discussed in \textcolor{red}{section \ref{sec:magnetic_field}}. Convection itself is not expected to drive the angular momentum transport at a level comparable to the MRI \citep[e.g.][]{Lesur2010, Held2018}, and in any case it is not self-sustainable, i.e., it requires an additional source of heat near disc midplane to establish the high temperature gradient.



\subsection{Energy transport by turbulent elements}
\textcolor{red}{In our model, we have not considered the possibility that the turbulent elements driving the angular momentum transport may also transport energy \citep{Ruediger1988}.} Such turbulent energy transport flux would be analogous to convection, transporting energy down the entropy gradient, the difference being that turbulent elements may persist at sub-adiabatic temperature gradients (as they are driven by other instabilities) in which case they transport energy from cooler to hotter regions \citep{Balbus2000}. We do not expect \textcolor{red}{including this mode of energy transport in our calculations to appreciably change} any of our results. \citet{DAlessio1998} found that turbulent energy transport accounts for less then 20\,\% of the total energy flux at any given orbital radius. In our work, the convectively stable upper layers of the disc are MRI-dead, and thus the thermal diffusivity due to turbulence is likely very low. In the regions of the disc where the radiative flux alone would yield super-adiabatic temperature gradient, we already assume that convection efficiently establishes the adiabat. The proportion of turbulent energy flux could be higher than the convective energy flux in such regions, but the temperature would not change significantly. \textcolor{red}{Hence, we have not included this effect in our analysis.}

\subsection{Ambipolar diffusion in the strong-coupling regime} \label{sec:ambipolar_diff_strong}
The criterion for ambipolar diffusion to quench the MRI that we employ is valid in the strong-coupling regime \citep{Bai2011a}. Strong coupling requires \textcolor{red}{that ionization equilibrium be} achieved on a timescale $t_{\rm chem}$ shorter than the dynamical timescale $t_{\rm dyn}=2\pi/\Omega$. Previously, \citet{Mohanty2018} reported that this condition is not fulfilled in most of the inner disc, as slow radiative recombinations make the chemical equilibrium timescale long. However, even in the absence of dust, three-body recombination (recombinations through collisions with the abundant molecular hydrogen) are much faster than radiative recombinations, and adsorption onto grains is even faster \citep{Desch2015}.

\begin{figure}
    \centering
    \includegraphics[width=0.49\textwidth,trim={0 0 0 0}, clip]{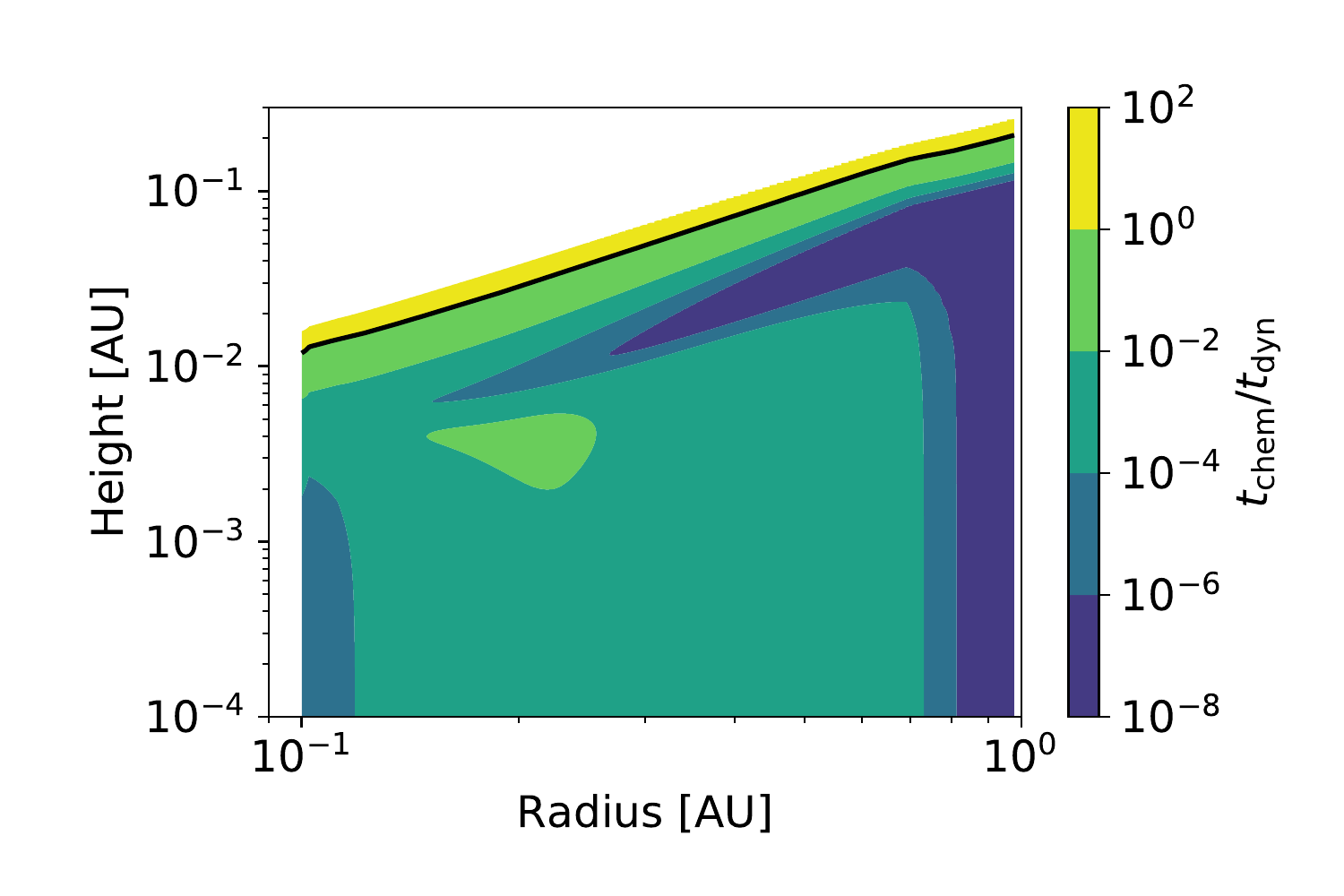}
    \caption{Ratio of the shortest recombination timescale ($t_{\rm chem}$) to the dynamical timescale ($t_{\rm dyn}$) as a function of the location in the disc. The solid line indicates where $t_{\rm chem}/t_{\rm dyn}=1$. Everywhere below this line the chemical equilibrium timescale is shorter than the dynamical timescale, justifying our assumption of the strong-coupling regime. See Section \ref{sec:ambipolar_diff_strong}.}
    \label{fig:chem_timescale}
\end{figure}

Since we solve directly for the equilibrium ionization state, we do not have access to the timescale $t_{\rm chem}$. However, we can estimate it as $t_{\rm chem}=n_{\rm e}/\mathcal{R}$, where $\mathcal{R}$ is the fastest of the above three recombination rates (in general, but not always, that is adsorption onto dust grains). Fig. \ref{fig:chem_timescale} shows, for our fully self-consistent model with our full chemical network, that $t_{\rm chem}/t_{\rm dyn} < 1$ everywhere except in the uppermost, lowest-density layers of the disc. We thus conclude that our use of the ambipolar diffusion criterion in the strong-coupling regime is justified.

\section{\textcolor{red}{Conclusions}} \label{sec:summary}
We present a steady-state model of the inner protoplanetary disc which accretes viscously, primarily due to the MRI. In this model, the disc is heated by viscous dissipation and stellar irradiation, and cools radiatively and convectively. The disc is ionized by thermal ionization, thermionic and ion emission from dust grains and by stellar X-rays, cosmic rays and radionuclides, and we also account for adsorption of charges onto dust grains. The disc's structure (density, temperature), viscosity due to the MRI, opacity and ionization state are calculated self-consistently everywhere in the disc (both as a function of radius and height). \textcolor{red}{To \textcolor{red}{the best of} our knowledge this is the first model that self-consistently couples all these processes to describe the structure of the inner regions of a steadily accreting protoplanetary disc.}

\textcolor{red}{We investigate how these various processes affect the structure of the inner disc and the extent to which the MRI can drive efficient accretion, i.e., the locations of the inner edge of the dead zone and the gas pressure maximum. For the fiducial parameters considered in this work \textcolor{red}{(stellar parameters: $M_*=1$\,M$_\odot$, $R_*=3$\,R$_\odot$, $T_*=4400$\,K, X-ray luminosity $L_{\rm X} = 10^{-3.5} L_{\rm bol}$; disk parameters: gas accretion rate $\dot{M}=10^{-8}$\,M$_\odot$\,yr$^{-1}$, viscosity in the MRI-dead zone $\alpha_{\rm DZ}=10^{-4}$; dust parameters: dust-to-gas mass ratio $f_{\rm dg} = 10^{-2}$, maximum grain size $a_{\rm max} = 1$\,$\mu$m)}, we find that:
\begin{enumerate}
    \item Inwards of the pressure maximum, the MRI is active \textcolor{red}{only} around the disc midplane. This \textcolor{red}{differs} from the predictions of vertically-isothermal models, and \textcolor{red}{is} possibly important for the evolution of dust grains in the inner disc. 
    \item Since the inner disc is optically thick, stellar irradiation \textcolor{red}{only weakly influences} the \textcolor{red}{midplane temperature, and thus also only} weakly affects the location of the dead zone inner edge.
    \item \textcolor{red}{Most of the inner disc is} convectively unstable, which we show is a property of any optically thick disc in which viscous heating \textcolor{red}{occurs} near the midplane. This motivates further work \textcolor{red}{into} a coupled MRI-convective instability in the limit of non-ideal MHD. 
    \item As suggested by \citet{Desch2015}, dust controls the ionization state of the inner disc, and thus the onset of the MRI. Thermal ionization plays a secondary role, as thermionic and ion emission from dust grains ionize the hot dense regions.
    \item High above disc midplane stellar X-rays produce an MRI-active layer. However, the X-rays barely change the overall viscosity at short orbital distances, or the location of the pressure maximum.
    \item \textcolor{red}{The pressure maximum resides at $\sim$0.7\,AU for our fiducial parameters, roughly the same location} as in our previous work \citep{Mohanty2018}. This is a consequence of the high optical depth in the inner disc, and the similarity between the ionization potential of potassium and the work function of the dust grains, rather than physics setting the pressure maximum location being similar.
\end{enumerate} }


\textcolor{red}{These conclusions are drawn for a disc with a fiducial dust-to-gas ratio of $10^{-2}$ and small dust grains ($a_{\rm max}=1$\,$\mu$m), which may be expected \textcolor{red}{in the early stages of dust evolution in the disc}. How these results depend on the model parameters, including dust grain size and dust-to-gas ratio, is explored in a companion paper (Jankovic et al. {\it in prep.}), where we \textcolor{red}{also} use our new inner disc model to speculate on \textcolor{red}{possible formation pathways for close-in super-Earths}.} Finally, it is important to note that our model is based on a simplifying assumption that the inner disc is at steady-state. Only time-dependent simulations will show whether steady-state is indeed achieved and, if it is, whether this steady-state is stable.

\section*{Acknowledgements}
\textcolor{red}{We thank the reviewer for helpful suggestions that improved the manuscript.} We thank Richard Booth, Thomas Haworth, Zhaohuan Zhu, Steven Desch, Neal Turner and Colin McNally for helpful discussions. M.R.J. acknowledges support from the President's PhD scholarship of the Imperial College London\textcolor{red}{ and the UK Science and Technology research Council (STFC) via the consolidated grant ST/S000623/1}. JEO is supported by a Royal Society University Research Fellowship. This project has received funding from the European Research Council (ERC) under the European Union’s Horizon 2020 research and innovation programme (Grant agreement No. 853022, ERC-STG-2019 grant, PEVAP). JCT acknowledges support from NASA ATP grant Inside-Out Planet Formation (80NSSC19K0010).

\section*{Data Availability}
The data underlying this article will be shared on reasonable request to the corresponding author.




\bibliographystyle{mnras}
\bibliography{bibliography} 



\appendix


\bsp	
\label{lastpage}
\end{document}